\RequirePackage{ifpdf}
\documentclass{JHEP3}

\usepackage{amsmath}
\usepackage{amssymb}
\usepackage{cancel}
\usepackage{slashed}
\usepackage{graphicx} 

\newcommand{\bea}{\begin{eqnarray}}
\newcommand{\eea}{\end{eqnarray}}
\newcommand{\beq}{\begin{equation}}
\newcommand{\eeq}{\end{equation}}
\newcommand{\bi} {\begin{itemize}}
\newcommand{\ei} {\end{itemize}}
\def\half{{\frac{1}{2}}}
\def\nl{\nonumber \\}
\def\br{\overline}
\def\wt{\widetilde}
\def\d{\partial}
\def\n{\nabla}
\def\LL {\mathcal {L}}
\def\JJ {\mathcal J}
\def\AA {\mathcal A}
\def\WW {\mathcal W}
\def\OO {\mathcal {O}}
\def\Dw{\Delta}
\def\l{\left(}
\def\r{\right)}
\def\gf{{ \bar g}}
\def\tabspace{\vrule height 3.5ex depth 2.2ex width 0pt}
\newcommand{\TO}[1] {\langle  \mathbf {T} \left \{#1\right \} \rangle}

\title{
The local Callan-Symanzik equation: structure and applications}

\author{
Florent Baume${}^{1,2}$, Boaz Keren-Zur${}^1$, Riccardo Rattazzi${}^1$ and Lorenzo Vitale${}^1$\\
${}^1$Institut de Th\'eorie des Ph\'enom\`enes Physiques, EPFL,\\
CH-1015 Lausanne, Switzerland\\
${}^2$Institut f\"ur Theoretische Physik, Ruprecht-Karls-Universit\"at, \\
 D-69120 Heidelberg, Germany\\
E-mails:  \email{ f.baume@thphys.uni-heidelberg.de, boaz.kerenzur@epfl.ch, riccardo.rattazzi@epfl.ch, lorenzo.vitale@epfl.ch}
}

\abstract{
The local Callan-Symanzik equation describes the response of a quantum field theory to local scale transformations in the presence of
background sources. The consistency conditions associated with this anomalous equation imply non-trivial relations among the $\beta$-function, the anomalous dimensions of composite operators and the short distance singularities of correlators.
In this paper we discuss various aspects of the local Callan-Symanzik equation and present new results regarding the structure of its anomaly.
We then use the equation to systematically write the n-point correlators involving the trace of the energy-momentum tensor. 
We use the latter result  to give a fully detailed  proof that the UV and IR asymptotics in a neighbourhood of a 4D  CFT must also correspond to  CFTs. 
We also clarify the relation between the matrix  entering the  gradient flow formula for  the $\beta$-function and a manifestly positive metric in coupling space associated with  matrix elements of the trace of the energy momentum tensor.
}


\begin{document}

 \section{Introduction}

The source method is a well established tool for probing the structure of Quantum Field Theory (QFT). The basic idea   is to promote the Lagrangian parameters (coupling constants and masses) to local background fields and to exploit the resulting 
(possibly local) symmetries to constrain the form of the effective action.
Moreover, the use of local sources allows to control the correlators of the associated composite operators, and, in particular, allows to  map the behavior of some operators across strongly coupled regimes.
 Prominent examples of the use of the source method are given
by the chiral lagrangian of  low-energy  hadrodynamics \cite{Gasser:1983yg} and by   exact results for holomorphic quantities in supersymmetric gauge theories \cite{Seiberg:1994bp}.  Another  playground where to usefully apply the method is given by softly broken supersymmetry, in perturbation theory \cite{Girardello:1981wz} and beyond \cite{ArkaniHamed:1998wc}. 

A crucial aspect of any given QFT  is  its behavior under renormalization group (RG) evolution. Technically, RG evolution corresponds to the change of  the dynamics under a dilation. In view of that, it   seems natural, in order to try and  explore the structure of the RG flow, to formally promote the explicitly 
broken dilation invariance to an exact Weyl symmetry. Of course, in order to be able to do that, one must promote the Lagrangian parameters to local fields with definite transformation property under Weyl symmetry. In particular the flat Minkowski metric $\eta_{\mu\nu}$ must  be upgraded to a generic curved metric  $g_{\mu\nu}$.
This program was carried out to a very significant extent    about two decades ago in a series of interesting papers by
Jack and Osborn \cite{Osborn:1989td,Jack:1990eb,Osborn:1991gm}. One first basic result is that the Weyl variation of the quantum effective action ${\WW}$  in the presence of sources is given by an anomaly equation\footnote{An earlier version of this equation was introduced already in 1979 by Drummond and Shore \cite{Drummond:1977dg}.}  
 \bea
 \label{eq_local_CS_intro}
\l 2g^{\mu\nu}\frac{\delta}{\delta g^{\mu\nu}(x)}-\beta^I(\lambda) \frac{\delta}{\delta\lambda^I(x)}+\ldots \r \WW [g,\lambda,\ldots]= \AA(x)
\eea
where $\lambda^I$ are the external sources, and ${\cal A}$ is a  local scalar function of these sources and the metric. In the case of a conformal field theory (CFT), by turning off all the sources apart from the metric, ${\cal A}$ reduces to the well known expression for the Weyl anomaly \cite{Capper:1974ic}. 
On the other hand, away from  criticality, where $\beta\not = 0$, this equation can be interpreted as a local generalization of the  Callan-Symanzik (CS) equation. Now, a second, perhaps more interesting set of results follows from the request of integrability of ${\cal A}$. This request can be enforced along two equivalent routes. One is to directly derive ${\cal A}$ from the bare lagrangian in a given renormalization scheme, 
for instance dimensional regularization \cite{Jack:1990eb}. The other is to require  ${\cal A}$ satisfies a Wess-Zumino consistency condition,  regardless of details concerning the renormalization scheme \cite{Osborn:1991gm}. The result 
is a set of non-trivial constraints  involving the $\beta$-functions and  the anomaly coefficients. The latter  can also  be interpreted as  the short distance singularities in different correlators involving  the energy momentum tensor and composite scalars and vectors. It is indeed according to that interpretation that some of these results had earlier  been
derived in works by Brown and Collins \cite{Brown:1980qq} and by Hathrell \cite{Hathrell:1981zb}. However, concerning 4D QFT, the most remarkable result of refs.  \cite{Jack:1990eb,Osborn:1991gm} is a relation involving the $\beta$-function and a quantity $\tilde a$ that coincides with the anomaly coefficient $a$ at critical points\footnote{$a$ is the coefficient of the Euler density term in the Weyl anomaly in 4 dimensions.}
\bea
\label{eq_CC_intro}
 \frac{\partial \tilde a}{\partial \lambda^I} =(\chi_{IJ}+\xi_{IJ})\beta^J
\eea
where $\chi$ and $\xi$ are respectively symmetric and antisymmetric covariant tensors over the space of couplings. Indeed, in the '70's,  a relation of this form had  been proved at finite loop order, and for specific models,  through a  laborious diagrammatic analysis \cite{Wallace:1974dy}. However  the use of the local CS  equation offers both a deeper viewpoint and a more systematic approach. Moreover, as $\tilde a$ only depends on the RG scale via its dependence  on the running couplings,  a corollary of the eq.~(\ref{eq_CC_intro}) is
\bea 
\mu\frac{d\tilde a}{d\mu}= \beta^I  \frac{\partial \tilde a}{\partial \lambda^I}=\chi_{IJ}\beta^J\beta^I\, .
\label{localatheorem}
\eea
This equation is fully analogous to the perturbative incarnation of Zamolodchikov's $c$-theorem \cite{Zamol} for 2D QFT, with $\chi_{IJ}$  interpreted as a {\it metric} in the space of couplings. 
Indeed the $c$-theorem itself can be shown to coincide with the  Wess-Zumino  consistency condition associated with  the 2D anomaly
off-criticality. More precisely, in the 2D case, as proven in ref. \cite{Osborn:1991gm},  there exists a choice of scheme where a quantity $\tilde c$, coinciding with $c$ at criticality, evolves according to the analogue of eq.~(\ref{localatheorem}), with a positive definite  metric. Concerning  the 4D case, although  in ref.  \cite{Jack:1990eb,Osborn:1991gm}   the positivity of $\chi_{IJ}$  could be established at leading order in perturbation theory, a robust non perturbative picture was missing. Perhaps because of this obstacle, no attempt to draw conclusions on the structure of 4D flows, in particular on their irreversibility, was made in those works.

Even in the absence of a proof, eq.~(\ref{localatheorem}), Cardy's conjecture \cite{Cardy:1988cwa} and direct evidence from exact results in supersymmetric gauge theories\cite{Anselmi:1997ys} had led to the belief that an irreversibility  argument for $a$, an $a$-theorem, should have existed in  the 4D case as well. But a  complete proof only arrived in 2011, in the work of Komargodski and Schwimmer (KS) \cite{Komargodski:2011vj,Komargodski:2011xv}, who showed that, in any flow between two CFTs, the end points of the flow satisfy the inequality $a_{UV}>a_{IR}$, where $a_{UV}$ ($a_{IR}$) is the value of the $a$ coefficient in the UV (IR) fixed point. With the wisdom of hindsight, it is now rather clear why the 4D proof took so much longer:
while for the c-theorem in 2D it suffices to  study the 2-point function of $T_{\mu\nu}$, the 4D analogue requires a  study of higher point correlators. This necessity had  already been noticed by Osborn \cite{Osborn:1991gm}, but within the local CS methodology there was no concrete guideline onto how to proceed. KS instead found a guideline in the form of an external background dilaton field,  the component of the background metric that   couples  to the trace $T$ of the energy momentum tensor. The  on-shell dilaton scattering amplitude  just happens to package the right combination of 2-, 3- and 4-point functions of $T$ that is directly sensitive to the RG flow of the anomaly coefficient $a$.
Using a dispersion relation for the scattering amplitude and using unitarity, KS could then compare the value of $a$ at the UV and IR asymptotics  and prove  $a_{UV}>a_{IR}$.
 
The $a$-theorem represents a non-perturbative constraints on the RG flow under the assumption that the end points are described by conformal field theories. However the same 
methodology introduced by KS gives a guideline to obtain further constraints on the structure of the flow, very much like it happens in 2D. A further step in this direction was given in ref.
 \cite{Luty:2012ww}, where the finiteness of the amplitude was used to exclude anomalous asymptotic behaviors for perturbative RG flows \footnote{That result was confirmed by an explicit study in weakly coupled gauge theories in ref. \cite{Fortin:2012hn}. As concerns ruling out anomalous asymptotics beyond perturbation theory, the specific case of 
 scale invariant field theory without conformal invariance was  cornered in ref. \cite{Luty:2012ww} and even more significantly so in ref. \cite{Dymarsky:2013pqa}. A totally clearcut proof is in our opinion still waiting, but probably imminent \cite{markus}.  }.
In a sense, the ingredients for this proof already existed in \cite{Jack:1990eb,Osborn:1991gm}, but the usage of the dilaton amplitude and dispersion relations made the connection to the asymptotics of the theory more transparent. Ref.  \cite{Luty:2012ww} provided a synthetic derivation relying on the minimal set of ingredients needed in a perturbative computation.  In particular, no detailed discussion of the structure and the role  of multiple insertions of $T$ was given. Moreover, issues like scheme dependence, operator mixing and the role of explicitly broken global symmetries were not  analyzed in full detail. Similarly the connection between the dilaton amplitude trick and eq.~(\ref{localatheorem}) was not fully explored.

The original goal of this paper was to illustrate all these details and to present a systematic method for computing correlation functions of $T$ off-criticality. 
We achieved this goal by studying and applying  the local Callan-Symanzik equation. 
 A by-product of this study is a new understanding of the structure of the Weyl anomaly. In practice we have shown that the anomaly can be written in a manifestly consistent manner up to the very  few terms related to the $a$ coefficient.

The first part of this paper, section 2, consists of a detailed analysis of the local Callan-Symanzik equation and is largely based on the original work by Osborn \cite{Osborn:1991gm}.
In particular, in section \ref{sec_setup} we present the equation and give a simple description of its derivation
(a more detailed discussion based on dimensional regularization is given in appendix \ref{app_CS_derivation}).
Section \ref{sec_local_CS} focusses on the generator of Weyl  transformations, and subtle issues involving its dependence on the scheme, choice of improvement and ambiguities in the presence of global symmetries. We also introduce new terminology and notations which are essential for the discussion in the following sections.
Next, in section \ref{sec_anomaly} we study the anomaly, which is  parameterized by  25 unknown tensor coefficients related by $\sim 10$ differential consistency conditions. We show that most of these conditions can be explicitly solved and  that the anomaly can be reformulated in a manifestly consistent form, with only 3 non-trivial consistency conditions remaining.
One combination of these is the famous equation \eqref{eq_CC_intro}, while two others,  involve anomalies related to external gauge fields. We then apply these results to the study of gradient flow formulae for the $\beta$-functions in section \ref{sec_gradient_flow}.

In the second part of the paper, we present a method for computing the $n$-point correlators of $T$, which we package  in terms of an effective dilaton action. 
We show how to express these correlators  as the sum of a local term related to the anomaly (section \ref{sec_local_interactions}) and  correlators of composite scalar operators (section \ref{sec_non_local}). Finally, in section \ref{sec_LPR} we use this machinery to revisit the results of ref. \cite{Luty:2012ww}. We also connect the dilaton based approach of  ref. \cite{Luty:2012ww} to the consistency condition approach of ref. \cite{Osborn:1991gm}.  As a by-product  we show that there exists a scheme where the metric $\chi_{IJ}$ essentially coincides with a  manifestly positive definite metric constructed in terms of combinations of matrix elements of composite operators. That is the analogue of what done in ref. \cite{Osborn:1991gm} for the 2D case. In section \ref{sec_conclusions}  we draw our conclusions.
   
 \section{The local Callan-Symanzik equation}
 \label{sec_CS}

 \subsection{General set-up}
\label{sec_setup}

\begin{figure}[htbp]
\begin{center}
		\includegraphics[height=120pt]{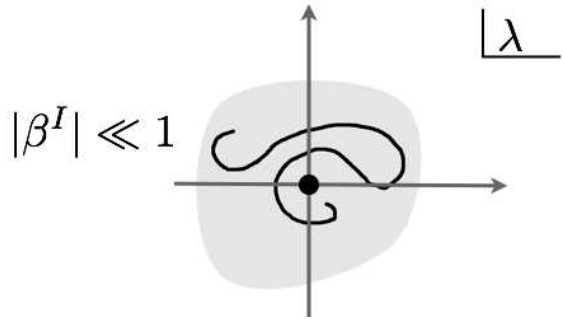}
	\caption{Our discussion concerns  RG flows in the vicinity of a conformal fixed point,  where the $\beta$-function and the anomalous dimensions can be treated as small perturbations.}
	\label{fig:flow}
\end{center}
\end{figure}

Our main goal is to study the properties of the RG flow in the neighbourhood of a conformally invariant fixed point.
The basic idea, as sketched in fig. \ref{fig:flow}, is to turn on all the possible marginal deformations of the CFT, which we describe  by a set of independent couplings $\lambda^I$, $I=1,\dots,N$, such that $\lambda^I=0$ corresponds to the unperturbed CFT. These couplings are associated with scalar operators ${\cal O}_I$, corresponding,  at the fixed point, to primaries with dimension equal to 4.  We shall moreover assume the original fixed point is endowed with an exact flavor symmetry $G_F$, which is in general explicitly broken at   $\lambda^I\not =0$.
One relevant question, originally addressed in ref. \cite{Luty:2012ww}, is to ask  which flows are possible and which are not, under the assumption that the asymptotics lie perturbatively  close to the original fixed point. An example  to which our assumption applies is given by weakly coupled renormalizable gauge theories with scalars and fermions. In that  case the original fixed point corresponds to free field theory. In particular it can be applied to the study of the flows in large N theories where one plays Banks-Zaks tricks \cite{Belavin:1974gu,BZ} to obtain novel fixed points or, possibly anomalous flows, such as SFTs (theories with scale but not conformal invariance) or limit cycles\footnote{As we already mentioned these exotic possibilities are now ruled out by the analysis in ref. \cite{Luty:2012ww}, which, among other things, we will here reproduce with extra details.}. However, our analysis also applies to the case where the original CFT represents a strongly coupled non-perturbative fixed point endowed with its own marginal deformations, like they are known to exist in supersymmetry. Indeed, as we shall be able argue later on,  our discussion applies  to the more general case in which there exists  an extended region of $\lambda$ space, where, even though the $\lambda^I$ may not be treated as small perturbations, the $\beta$-function can still be treated as small. Examples of this  more general case can be found in QFTs with  manifolds of fixed points (see for instance \cite{Leigh:1995ep}). While we do not know of any explicit  examples
in theories without supersymmetry, we believe consideration of this possibility, even if merely conceptual, better illustrates what are the necessary ingredients in our study.

In QFT the trace of the energy momentum tensor  $T\equiv T^\mu_\mu$ is known to correspond to the divergence of the naive dilation current.  The change of the dynamics under (naive) dilations is thus controlled by correlators involving $T$. In order to make the properties of these correlators more explicit, we need to expand $T$ in a complete basis of scalar operators of dimension 4. This basis surely includes the scalar deformations ${\cal O}_I$ that generate the flow, but in principle there could also appear divergences of the flavor currents $\partial_\mu J^\mu_A$ and operators of the form $\Box {\cal O}_a$ where, at the fixed point,  ${\cal O}_a$, are primary scalars of dimension 2. It is therefore crucial to have a convenient method to control the properties of these operators.  Now, the standard methodology  to define composite operators and their correlators is to introduce the associated spacetime dependent sources. For instance, the energy momentum tensor $T^{\mu\nu}$ will have as its source a local background metric $g_{\mu\nu}(x)$, while ${\cal O}_I$ will have as its source a spacetime dependent coupling $\lambda^I(x)$. Along the same line, in order to source the currents $J^\mu_A$, we shall turn on background vector fields $A_\mu^A(x)$ gauging the flavor group $G_F$, while   the dimension 2 operators ${\cal O}_a$ will be sourced by  scalar fields $m^a(x)$.
We shall collectively  indicate the  set of local sources by ${\cal J}\equiv (g^{\mu\nu},\lambda^I, A_\mu^A, m^a)$. The renormalized partition function in the source background 
\beq
Z[{\cal J}]\equiv e^{i {\cal W}[{\cal J}]}=\int {\cal {D}} \Phi e^{i S[\Phi, {\cal J}]}
\eeq
acts as the generator of the correlators for the associated renormalized composite operators. The same information is more efficiently encapsulated in the quantum effective action  ${\cal W}$, which generates the connected correlators. When acting on ${\cal W}$ the functional derivative with respect to a source coincides with the insertion of the corresponding operator in a connected correlator
\bea
\frac{2}{\sqrt{ - g}}\frac{\delta }{\delta g^{\mu\nu}(x)}\equiv \left [ T_{\mu\nu}(x)\right]
&\qquad&
\frac{1}{\sqrt{ - g}} \frac{\delta }{\delta \lambda^I(x)}\equiv\left [ \OO_I(x)\right]\nl
\frac{1}{\sqrt{ - g}}\frac{\delta }{\delta A_\mu^A(x)}\equiv \left [ J^A_\mu(x)\right]
 &\qquad&
\frac{1}{\sqrt{ -g}}\frac{\delta}{\delta m^a(x)}\equiv\left [\OO_a(x)\right]~.
\label{source-operator1}
\eea
Time ordered n-point correlators are obtained by first taking  n-derivatives of ${\cal W}$ and then setting the sources to ``zero": $g^{\mu\nu}(x)\to \eta^{\mu\nu}$, $\lambda^I(x)\to \lambda^I ={\rm {const}}$, $A_\mu^A=0$ and  $m^a(x)\to m^a ={\rm {const}}$. We will use the following convention:
\bea
\TO{\OO_{I_1}(x_1)\ldots \OO_{I_n}(x_n)}&=&
\frac{(-i)^{n-1}}{\sqrt {-g(x_1)}\ldots \sqrt{-g (x_n)}}~
\frac{\delta}{\delta \lambda^{I_n}(x_n)}\ldots\frac{\delta}{\delta \lambda^{I_1}(x_1)}\WW \Big|\nl
\TO {T(x_1)\ldots T(x_n)}&=&
\frac{(-i)^{n-1}2^n  }{\sqrt {-g(x_1)  }\ldots \sqrt{-g (x_n)}}~
g^{\mu_n\nu_n} \frac{\delta}{\delta g^{\mu_n\nu_n}(x_n)}\ldots g^{\mu_1\nu_1}\frac{\delta}{\delta g^{\mu_1\nu_1}(x_1)}\WW \Big|~.\nl
\label{nT}
\eea
Notice that our definition of the n-point correlator of $T$ coincides with the standard one
\beq
\TO{ T(x_1)\ldots T(x_n)}_{S}=\frac{(-i)^{n-1}2^n  }{\sqrt {-g(x_1)  }\ldots \sqrt{-g (x_n)}}~
g^{\mu_n\nu_n}\ldots g^{\mu_1\nu_1} \frac{\delta}{\delta g^{\mu_n\nu_n}(x_n)}\ldots\frac{\delta}{\delta g^{\mu_1\nu_1}(x_1)}\WW \Big|
\eeq
up to contact terms.

A standard property of effective actions for  sources is to formally respect extended symmetries, up to anomalies.  As concerns  diffeomorphisms and  $G_F$ transformations, in this paper we shall make the simplifying assumptions that they are anomaly free. 
Indeed most of our discussion shall focus on the case of parity  invariant theories, for which  ${\mathrm {diff}} \times G_F$ are not anomalous \footnote{We plan to get back to the general case of anomalous global symmetries in a forthcoming paper.}.
The other crucial symmetry  is given by Weyl transformations under which the metric transforms as
\beq
g^{\mu\nu}(x)\to e^{2\sigma(x)} g^{\mu\nu}(x)\qquad\qquad \delta_\sigma g^{\mu\nu}(x)=2\sigma(x) g^{\mu\nu}(x)~.
\eeq
and whose anomaly is the centerpiece of our study. The origin of the Weyl anomaly is discussed in more detail in the appendix, focussing on dimensional regularization. Here we shall limit ourselves to the basic story, which goes as follows.
As a function of  the sources ${\cal J}\equiv (g^{\mu\nu},\lambda^I, A_\mu^A, m^a)$ and of the dynamical fields the bare action  can be in general split as
\beq
S=S^{(1)}[\Phi, {\cal J}]+S^{(2)}[{\cal J}]\,
\label{s1+s2}
\eeq
 where $S^{(1)}$ involves only terms that non-trivially depend on the dynamical fields, while  $S^{(2)}$ contains, instead, purely source dependent terms such as $(\n^2\lambda)^2$, $R(\n\lambda)^2$, $R_{\mu\nu}R^{\mu\nu}$, etc.. The addition of $S^{(2)}$ is necessary in order to obtain a finite quantum effective action after renormalization. In dimensional regularization  $S^{(2)}$ can be chosen to be a series of pure poles in $1/\epsilon$. Now, given that ${\cal J}$ represent the complete set of sources for the operators that can appear in the expansion of $T$, it is basically by definition that there must exist a choice of Weyl transformation $\delta_\sigma{\cal J}$ such that $S^{(1)}$ is invariant. Once again, as we show in the appendix,  in dimensionally regulated weakly coupled gauge theories, this fact is pretty obvious. On the other end, once $\delta_\sigma {\cal J}$ is picked that way, it is clear that  $S^{(2)}$
 will in general not be invariant \footnote{Unless new sources, coupling to pure functions of ${\cal J}$ are introduced, in such such a way that their variation compensates for  $\delta_\sigma S^{(2)}$.}. Since $S^{(2)}$ has no dependence on the dynamical fields, its variation will directly control the variation of the quantum effective action. We thus have
 \beq
 \int d^4x \,\delta_\sigma {\cal J}\frac{\delta }{\delta {\cal J}} {\cal W} = \int d^4x\, \delta_\sigma {\cal J}\frac{\delta }{\delta {\cal J}} S^{(2)} \equiv \int d^4 x\,{\cal A_\sigma}
 \label{sketchCS}
\eeq
where the locality of $S^{(2)}$ dictates ${\cal A_\sigma}$  must  be a local  function of the sources. Notice moreover that, even though $S^{(2)}$ is a series of counterterms that diverge with the cut-off,  by eq.~(\ref{sketchCS}), its variation $\int {\cal A_\sigma}$ equals the variation of the renormalized action with respect to the renormalized sources, and must therefore be finite.  ${\cal A_\sigma}$  represents an anomaly for the  Weyl symmetry. Eq.~(\ref{sketchCS}) is the local Callan-Symanzik equation  we sketched in eq.~(\ref{eq_local_CS_intro}). 

 \subsection{The structure of  Weyl symmetry}
 \label{sec_local_CS}

In this section we analyze in detail the Weyl transformation of the sources.
The discussion is based mainly on \cite{Osborn:1991gm}, but we shall highlight properties which we repute relevant to the study of the anomaly and to the computation of the dilaton effective action\footnote{As further reading material we recommend \cite{Nakayama:2013ssa,Nakayama:2013wda}.}.

Let us recall once more the role of our sources.  The dimensionless sources $\lambda^I(x)$, associated with quasi marginal operators $\OO_I(x)$, are local versions of the couplings $\lambda^I$ that produce the RG flow we want to study. The CFT fixed point we are expanding around corresponds to $\lambda^I=0$. This fixed point  respects a flavor symmetry $G_F$, which is in general explicitly broken at  $\lambda^I\not =0$. The vectors  $A_{\mu}^A$, with the  index $A$ running in the adjoint of $G_F$, are background  fields gauging $G_F$. They act as sources for the currents $J^\mu_A$.  By the  scalars $m^a(x)$, we indicate the sources of scalar operators ${\cal O}_a$ with dimension  equalling 2 at the fixed point. Notice that $m^a$ have mass dimension two, in spite of the perhaps misleading notation (which we adopted from ref. \cite{Osborn:1991gm}). The CFT may also possess relevant scalar deformations of dimension $\not = 2$. For instance, in weakly coupled gauge theories these are given by fermion masses and scalar trilinears, that are associated with dimension 3 operators. In the limit where the corresponding mass deformations vanish the appearance of these  operators  in the expansion of $T$ is forbidden by Lorentz invariance. We shall thus neglect them in the course of our discussion. Finally notice that, although we do not indicate it, the sources  and the corresponding composite operators in eq.~(\ref{source-operator1}) are defined at some renormalization scale $\mu$.

The discussion in this section is not affected by the assumption of parity conservation. As it will be clear from eq.~(\ref{deltasource}), that is simply because,  by dimensional analysis, the Levi-Civita tensor $\epsilon^{\mu\nu\rho\sigma}$ cannot appear in the Weyl transformation  of the sources. The situation is however different for  the Weyl anomaly discussed in section 2.3. Notice that for parity invariant theories,  $G_F$ should be thought as a  (maximal) vector subgroup of the full flavor group.

The Weyl symmetry generator is  the sum of the variations of the
complete set of sources ${\cal J}=(g^{\mu\nu},\lambda^I,A_\mu^A,m^a)$
\beq
\delta_\sigma {\cal J}\frac{\delta }{\delta {\cal J}}\equiv \Delta_\sigma =\Delta^g_\sigma-\Delta^{\beta}_\sigma
\eeq
where
\bea
\label{eq_local_CS_definition}
\Delta^g_\sigma&=& \int d^4x~2\sigma g^{\mu\nu}\frac{\delta}{\delta g^{\mu\nu}}\nl
\Delta^\beta_\sigma &=&- \int d^4x \left (\delta_\sigma\lambda\cdot \frac{\delta}{\delta\lambda}+\delta_\sigma A_\mu\cdot \frac{\delta}{\delta A_\mu}+\delta_\sigma m\cdot \frac{\delta}{\delta m}\right )\,.
\eea
The Weyl variation of the sources will have the most general form compatible with dimensional analysis (power counting) and symmetry (diffeomorphisms and $G_F$). That is:
\bea
\delta_\sigma\lambda^I &=&-\sigma \beta^I\nl
\delta_\sigma A_\mu^A&=& -\sigma  \rho_I^A  \n_\mu \lambda^I+ \n_\mu\sigma S^A\nl
\delta_\sigma m^a &=&\sigma\l m^b\l 2\delta_b^a  - \br \gamma_b^a\r 
+ C^a R+  D_I^a\n^2 \lambda^I + \half E_{IJ}^a\n_\mu \lambda^I \n^\mu \lambda ^J\r
- \n_\mu\sigma  \theta_I^a \n^\mu \lambda^I+ \n^2 \sigma  \eta^a\nl
\label{deltasource}
\eea
where  $\n$ to denotes the $G_F$  covariant derivative
\bea
\n_\mu\lambda^I = \d_\mu \lambda^I+ A^A_\mu (T_A\lambda)^I
\eea
and $T_A$ is a generator of $G_F$. By dimensional analysis, the various coefficients $\beta^I, \rho_I^A,\dots, \eta^a$ in eq.~(\ref{deltasource})  are  functions of the marginal couplings $\lambda^I$. Moreover, as the Weyl symmetry commutes with $G_F$, these coefficients should be covariant functions.
It would be straightforward to add to this set up the sources $\wt  m^\alpha$ of relevant scalar deformations having  dimension $\not =2$ at the original fixed point. By dimensional analysis the transformation would simply reduce to
\beq
\delta_\sigma \wt  m^\alpha =\sigma D_\beta^\alpha \wt  m^\beta
\eeq
with $D_\beta^\alpha$  a $\lambda$ dependent matrix whose eigenvalues differ from $2$ in the whole neighborhood of the fixed point we are studying. Notice that unlike for the case of $m^a$ in eq.~(\ref{deltasource}), the dimensionality of $\wt  m^\alpha$ forbids the presence  of terms involving $R(g)$ or derivatives of $\sigma$ and $\lambda$.
 
The local Callan-Symanzik can thus be written as
\bea
\label{eq_local_CS}
\Delta_\sigma \WW = ( \Delta^g _\sigma-\Delta^\beta _\sigma)  \WW &=&\int d^4 x\AA_\sigma~.
\eea
We shall now study  the Weyl generator $\Dw_\sigma$ in detail, 
focussing on properties that will help clarify the structure of the anomaly and also help compute the matrix elements of $T$.

\subsubsection{The global CS equation, dilations and conformal transformations}

It is important to relate the Weyl symmetry generator  $\Delta_\sigma$ to the other incarnations of dilations. First we must relate it to RG transformations, which are obtained as follows. Consider first  all the classically dimensionful parameters appearing in $\WW$. In our case these are just the renormalization scale $\mu$ and the dimension two sources $m^a$. Accounting for the fact that lengths are purely controlled by $g_{\mu\nu}$, we have then the obvious identity
\beq
\Delta^{\mathrm {\mu}} \,{\cal W}\equiv \left [\mu\frac{\partial}{\partial\mu}+\int d^4x \left(2m^a(x)\frac{\delta}{\delta m^a(x)}+2g^{\mu\nu}(x)\frac{\delta}{\delta g^{\mu\nu}(x)}\right)\right ]{\cal W}=0
\eeq
By combining the above operator with a Weyl generator with constant parameter $\sigma=-1$ in such a way as to eliminate the derivative with respect to the metric we obtain 
\beq
\Delta^{RG}\equiv \Delta^{\mathrm {\mu}}+\Delta_{\sigma=-1}= \mu\frac{\partial}{\partial\mu}+\int d^4x \left(\beta^I\frac{\delta}{\delta \lambda^I(x)}+\bar\gamma^a_bm^b(x)\frac{\delta}{\delta m^a(x)}+\dots\right)
\eeq
which corresponds to the ordinary Callan-Symanzik operator generalized to the case of local sources. The RG transformation of the effective action, $\Delta^{RG}{\cal W}$, is simply the integral of the  Weyl anomaly for constant $\sigma$. This result establishes  a direct connection between the terms in the anomaly and the explicit dependence on  $\ln\mu$ of ${\cal W}$. This dependence is associated with
logarithmic UV divergences. We shall further discuss this connection in section \ref{sec_anomaly}.

The other important incarnations  are  global dilations and special conformal transformations. They correspond to those particular combinations of a diffeomorphism and a Weyl transformation that leave the flat metric $\eta_{\mu\nu}$ invariant. The generator of infinitesimal diffeomorphisms is defined by
\bea
\Delta^{Diff}_\xi &=&\int d^4x \l \l \n_\rho \xi^\mu g^{\rho\nu}+\n_\rho \xi^\nu g^{\mu\rho}\r   \frac{\delta}{\delta g^{\mu\nu}}
- \n_\mu \xi^\nu A^A_\nu \frac{\delta}{\delta A^A_\mu} \r\nl
&&-\int d^4x \xi^\rho \l 
 \n_\rho\lambda^I \frac{\delta}{\delta \lambda^I}
+ \n_\rho A^A_\nu \frac{\delta}{\delta A^A_\nu}
+ \n_\rho m^a \frac{\delta}{\delta m^a}
\r\, .
\eea
Our assumption that diffeomorphism are non-anomalous corresponds to $\Delta^{Diff}_\xi {\cal W}=0$ for any $\xi$. 
An infinitesimal dilation  is given by the following combination of a diffeomorphism and a Weyl transformation \bea
 \xi^\mu=cx^\mu\qquad \qquad \sigma = -c
\eea
The corresponding generator  is
\bea
\Delta^D_c 
&\equiv&\Delta^{Diff}_{\xi= cx} + \Delta_{\sigma=-c}\nl
&=&
c \int d^4x\l\beta^I \frac{\delta}{\delta \lambda^I}+    \l   \rho_I^A  \n_\mu \lambda^I -A^A_\mu \r \frac{\delta}{\delta A^A_\mu}\r \nl
\nl
&& -c \int d^4x\l m^b\l 2\delta_b^a  - \br \gamma_b^a\r 
+C^a R+  D_I^a\n^2 \lambda^I + \half E_{IJ}^a\n_\mu \lambda^I \n^\mu \lambda ^J\r\frac{\delta}{\delta m^a} \nl
&&-c\int d^4x~ x^\rho \l 
 \n_\rho\lambda^I \frac{\delta}{\delta \lambda^I}
+ \n_\rho A^A_\nu \frac{\delta}{\delta A^A_\nu}
+ \n_\rho m^a \frac{\delta}{\delta m^a}
\r
\eea
Infinitesimal special conformal transformations are instead given by 
\bea
\xi^\mu=2(b\cdot x) x^\mu - x^2 b^\mu \qquad \qquad \sigma = -2b\cdot x
\eea
so that the corresponding generator is 
\bea
\Delta^K_{b} &\equiv&\Delta^{Diff}_{\xi= \l 2(b\cdot x) x^\mu - x^2 b^\mu\r } + \Delta_{\sigma=-2b\cdot x}\nl 
&=&
2b_\mu \int d^4x\l x^\mu  \beta^I \frac{\delta}{\delta \lambda^I}+  \l x^\mu \l   \rho_I^A  \n_\nu \lambda^I- A^A_\nu  \r  -\delta^\mu_\nu S^A   \r \frac{\delta}{\delta A^A_\nu}  \r \nl
&& -2b_\mu \int d^4x\l x^\mu \l m^b\l 2\delta_b^a  - \br \gamma_b^a\r 
+ C^a R+  D_I^a\n^2 \lambda^I + \half E_{IJ}^a\n_\mu \lambda^I \n^\mu \lambda ^J\r-  \theta_I^a \n^\mu \lambda^I\r \frac{\delta}{\delta m^a}  \nl
&&-\int d^4x \l 2(b\cdot x) x^\rho - x^2 b^\rho\r \l 
 \n_\rho\lambda^I \frac{\delta}{\delta \lambda^I}
+ \n_\rho A^A_\nu \frac{\delta}{\delta A^A_\nu}
+ \n_\rho m^a \frac{\delta}{\delta m^a}
\r
\label{Kgenerator}
\eea

QFTs that are invariant under dilations (and conformal tranformations) correspond to points in source space that are left invariant by the action of $ \Delta^D$ (and $\Delta^K$).
As expected, a point $\lambda^I=\lambda^I_*=const$, such that $\beta^I=0$, with also $g_{\mu\nu}=\eta_{\mu\nu}$, $A_\mu^A=m^a=0$ satisfies dilation invariance. On the other hand, from the explicit form of $\Delta^K$, one sees that the condition for conformal invariance is a different one. In particular, if $\beta=0$ while $S^A\not =0$ we have an SFT, that is a QFT with scale invariance but without conformal invariance.

\subsubsection{The local CS equation and the operator algebra}

Equation \eqref{eq_local_CS} encapsulates the relation between  $T$ and the other composite operators. 
By iterating  the equation we find this relation for any number of insertions of $T$. We can consider the following distinct cases:
\begin{itemize}
\item When none of the points in the time ordered correlator coincide, then by eq.~\eqref{eq_local_CS}  we can write
\bea
\TO{  T(x) \ldots} &\supset&\beta^I \TO{ \OO_I(x)\ldots } + S^A\TO{ \n_\mu J_A^\mu(x)\ldots } - \eta^a  \TO{ \n^2 \OO_a(x)\ldots}\nl
\eea
This can be understood as an operator equation for $T$:
\bea
\label{eq_T_beta_S_t}
 T &=&\beta^I \left[\OO_I\right] + S^A \n_\mu  \left [J_A^\mu\right] - \eta^a  \n^2  \left [\OO_a\right]~.
\eea
The coefficients $\beta^I$, $S^A$ and $-\eta^a$ are the coordinates of $T$ in the space of dimension 4 composite operators.

\item When two, or more, points coincide, we find contact terms proportional to variations of the  coefficients in the Weyl generator, e.g.
\bea
\TO{  T(x) \OO_I(y) \ldots} &\supset&-i \delta(x-y)\Big( \d_I\beta^J \TO{ \OO_J(x)\ldots } - \rho_I^A\TO{  \n_\mu J_A^\mu(x)\ldots } \nl
&&~~~~~~~~~~~~~~~- D_I^a  \TO{ \n^2 \OO_a(x)\ldots }\Big) \nl
\TO{  T(x) \OO_I(y) \OO_J(z)\ldots} &\supset&-\delta(x-y)\delta(x-z) E^a_{IJ}  \TO{ \OO_a(x)\ldots }
\eea
\item When all points coincide, there are additional  ultra-local contributions encoded by  the 
Weyl anomaly. These will be   discussed in section \ref{sec_anomaly}.

\end{itemize}

It is also interesting to consider the field operator interpretation of the commutators of the source differential operators with $\Delta^{RG}$, $\Delta^D$ and $\Delta^K$ defined in the previous section. In particular the commutators with $\Delta^{RG}$ control the renormalization scale dependence of the corresponding renormalized composite operators. For instance we have
\beq
[\Delta^{RG}, \frac{\delta}{\delta\lambda^I(x)}]= -\partial_I\beta^J \frac{\delta}{\delta\lambda^J(x)}+\dots\qquad\rightarrow\qquad \mu\frac{d}{d \mu} {\cal O}_I=-\partial_I\beta^J{\cal O}_J+\dots
\eeq
The commutators with $\Delta^D$ and $\Delta^K$ control the transformation of the  composite operators in the Ward identities for the corresponding (generally explicitly broken) symmetries. At the special symmetry preserving points in parameter space these can be interpreted as the commutator with the corresponding conserved charges $D$ and $K^\mu$. The explicit computation of the commutators among the various functional differential operators leads to the following results
\begin{equation}
\mu \frac{d}{d \mu} 
\begin{pmatrix}
T \\ O_a \\ J^\mu_A \\ O_I
\end{pmatrix} =
\left(
\begin{matrix}
0 &  6 C^b \nabla^2 & 0 & 0 \\
0 & - \bar{\gamma}_a^{~b} & 0 & 0 \\
0 & D^b_K (T_A \lambda)^K\n^\mu~~ & - \rho^B_K (T_A \lambda)^K & 0 \\
0 & D^b_I \nabla^2 & \rho^B_I \nabla_\mu & -\partial_I \beta^J \\
\end{matrix}
\right) 
\begin{pmatrix}
T \\ O_b \\ J^\mu_B \\ O_J
\end{pmatrix}
\label{RGflowoperators}
\end{equation}

\begin{equation}
\label{eq_dilations}
D 
\begin{pmatrix}
T \\ O_a \\ J^\mu_A \\ O_I
\end{pmatrix} =
\left(
\begin{matrix}
4 & - 6 C^b \nabla^2 & 0 & 0 \\
0 & 2 \delta_a^{~b} + \bar{\gamma}_a^{~b} & 0 & 0 \\
0 & -D^b_K (T_A \lambda)^K\n^\mu~~ & 3 \delta_A^{~B}+ \rho^B_K (T_A \lambda)^K & 0 \\
0 & - D^b_I \nabla^2 & -\rho^B_I \nabla_\mu &4\delta_I^J +  \partial_I \beta^J \\
\end{matrix}
\right) 
\begin{pmatrix}
T \\ O_b \\ J^\mu_B \\ O_J
\end{pmatrix}
\end{equation}

\begin{equation}
\label{eq_K_transformations}
K^\mu 
\begin{pmatrix}
T \\ O_a \\ J^\nu_A \\ O_I
\end{pmatrix} = 2
\left(
\begin{matrix}
 0 &  6 C^b \nabla^\mu & 0 & 0 \\
0 & 0 & 0 & 0 \\
0 & - (D^b_K + \theta^b_K)(T_A \lambda)^K g^{\mu \nu} & 0 & 0 \\
0 & (2 D^b_I + \theta^b_I) \nabla^\mu &  \rho^B_I + \partial_I S^B  & 0\\
\end{matrix}
\right) 
\begin{pmatrix}
T \\ O_b \\ J^\mu_B \\ O_J
\end{pmatrix}
\end{equation}
Focussing on  fixed points, we shall later comment on the consistency of the above results with the algebra of unitary conformal field theory.

\subsubsection{Ward identities and ambiguities}
\label{sec_ambiguity}
The basis of renormalized operators used to write $T$ in eq.~(\ref{eq_T_beta_S_t}) is redundant in the presence of symmetries.
Indeed, by the equations of motion, $\n_\mu  J_A^\mu$ equals the $G_F$ variation of the Lagrangian and can thus be expressed in terms of
a combination of ${\cal O}_I$ and ${\cal O}_a$.
In the background source approach  this  is viewed by  considering
 the $G_F$  Ward identity ($\alpha^A(x)$ are the  Lie parameters of $G_F$)
\bea
\Delta^F_{\alpha} \WW&\equiv&
\int d^4x\Bigg[ \alpha^A \l (   T_A \lambda)^I  \frac{\delta}{\delta \lambda^I(x)}+ (  T_A m) ^a \frac{\delta}{\delta m^a(x)}\r 
- \n_\mu \alpha^A  \l  \frac{\delta}{\delta A^A_\mu(x)}\r\Bigg]  \WW =0\nl
\label{eq_WI}
\eea
which simply translates  into the operator equation
\beq
\label{eq_naive_WI}
(T_A\lambda)^I \OO_I+(T_A m)^a \OO_a+\n_\mu  J_A^\mu=0\, .
\eeq

An alternative procedure is to define a new Weyl  generator by combining  the original $\Delta_\sigma$  with an infinitesimal   $G_F$ transformation with Lie parameter $\alpha^A(x)=-\sigma(x)\omega^A(\lambda)$ 
\beq
\Delta_\sigma \to \Delta'_\sigma \equiv \Delta_\sigma +\Delta^F_{-\sigma \omega}
\label{flavorredef}
\eeq
Provided $\omega^A(\lambda)$  is chosen to be a covariant (but otherwise arbitrary) function of the $\lambda$'s, the redefined Weyl symmetry  still commutes with $G_F$. Eq.~(\ref{flavorredef}) corresponds to the following redefinition of the coefficients of the local CS operator:  
\bea
\begin{array}{lllllll}
\label{eq_operator_gauge_freedom}
\beta^I&\to& \beta^I+\l \omega^A T_A \lambda\r^I&\qquad&
\br\gamma_b^a&\to& \br\gamma_b^a + \l \omega^A T_A\r_b^a\\
S^A&\to& S^A+\omega^A&\qquad&
\rho_I^A&\to& \rho_I^A-\d_I\omega^A~.
\end{array}
\eea
Notice that this is an ambiguity  inherent in the definition of the $\beta$-function and of the anomalous dimensions \cite{Osborn:1991gm,Tseytlin:1986tt}. When carrying out the renormalization procedure this ambiguity corresponds to the freedom in defining the wave function renormalization matrix relating bare and renormalized fields \cite{Luty:2012ww}.

The  redundancy in  the definition of  $\Delta_\sigma$  is quite analogous to   a gauge symmetry.  Like for gauge symmetry, unambiguous physical information is carried by  the invariants, which in our case these are given by  
\bea
\label{eq_def_B}
B^I&=& \beta^I-\l S^AT_A\lambda\r^I\nl
 P_I^A&=&\rho_I^A +\d_I S^A\nl
 \gamma_b^a&=& \br \gamma_b^a- \l S^AT_A\r_b^a\, .
 \eea
 These are the quantities that unambiguously describe the RG flow. Indeed they correspond to fixing the ``gauge" by choosing $\omega^A=-S^A$ in eq.~(\ref{eq_operator_gauge_freedom}) so that the redefined $S^A$ vanishes. 
Correspondingly, by solving for $\n_\mu  J_A^\mu$ in eq.~(\ref{eq_naive_WI}), at $m_a=0$,    $T$ in eq.~(\ref{eq_T_beta_S_t}) reads
\bea
 T &=&B^I \left[\OO_I\right] -\eta^a  \n^2  \left [\OO_a\right]~.
 \label{TB}
\eea

Notice that  by the change in eq.~(\ref{flavorredef}) also the  $\Delta^{RG}$ acquires an extra flavor rotation term. Making the choice $\omega^A=-S^A$ and using eq.~\eqref{eq_naive_WI}, the RG transformation of the renormalized operators becomes  then (disregarding the contribution from ${\cal O}_a$)
\begin{equation}
\mu \frac{d}{d \mu} 
\begin{pmatrix}
 J^\mu_A \\ O_I
\end{pmatrix} =
\left(
\begin{matrix}
- P^B_K (T_A \lambda)^K & 0 \\
 0   & -\l  \d_IB^J+P_I^C(T_C\lambda)^J\r \\
\end{matrix}
\right) 
\begin{pmatrix}
J^\mu_B \\ O_J\, .
\end{pmatrix}
\end{equation}
With this definition, we can identify the following matrices as the anomalous dimensions of the composite operators
\bea
\label{eq_anomalous_dimensions}
\gamma^J_I&=& \d_IB^J +P^A_I(T_A\lambda)^J\nl
\gamma^B_A&=& P^B_K(T_A\lambda)^K~.
\eea

\subsubsection{Lie derivatives}

A recurrent object that will appear in the analysis  is a variant of the Lie derivative, which describes the Weyl transformation of covariant tensors
\bea
\label{eq_def_Lie}
\LL [Y_{IAa\ldots}^{JBb\ldots}]&=& B^K\d_K Y_{IAa}^{JBb} 
+\gamma_I^K Y_{KAa\ldots}^{JBb\ldots}
+\gamma_A^C Y_{ICa\ldots}^{JBb\ldots}
+\gamma_a^c Y_{IAc\ldots}^{JBb\ldots}\nl
&&~~~~~~~~~~~~~~~~
-\gamma_K^J Y_{IAa\ldots}^{KBb\ldots}
-\gamma_C^B Y_{IAa\ldots}^{JCb\ldots}
-\gamma_c^b Y_{IAa\ldots}^{JBc\ldots}+\ldots
\eea
where the different $\gamma$ matrices were defined in \eqref{eq_def_B} and \eqref{eq_anomalous_dimensions}. The operator $\LL$ so defined satisfies the distributive property of derivatives when considering products of tensors, including contractions of covariant and contravariant indices. Schematically one has
\beq
\LL [Y\cdot Z]=Y\cdot \LL[Z]+\LL[Y]\cdot Z\,.
\eeq
For instance one has $\LL [Y_A^I\cdot Z^A]=Y_A^I\LL[Z^A]+\LL[Y_A^I]Z^A$. Moreover one can easily check that the tensor $ v_A^I\equiv (T_A\lambda)^I$ satisfies $\LL[v_A^I]=0$ and can thus be carried freely in and out of the $\LL$ symbol. The latter  property  depends crucially on eq.~(\ref{eq_anomalous_dimensions}) which relates the anomalous dimensions for scalars and currents. The Lie derivative appears, for example,  in  the Weyl variation of space-time derivatives of the sources
\bea
\label{eq_var_nabla_lambda}
\Delta_\sigma \l Y_I \n^\mu\lambda^I\r &=& \sigma \l -\LL[ Y_I]  \n_\mu\lambda^I  \r +\n_\mu \sigma\l -B^I  Y_I\r
\nl
\Delta_\sigma \l Y_I  \n^2 \lambda^I\r &=&\sigma\l 2Y_I \n^2 \lambda^I -  \LL[Y_I ]  \n^2 \lambda^I - 
Y_I U^I_J ~\gamma^J_{KL}
\n_\mu \lambda^K\n^\mu\lambda^L\r \nl
&&~~~~~~~~+\n_\mu\sigma\l -2 Y_IU^I_J  \n^\mu \lambda^J\r  +   \n^2\sigma \l - B^I Y_I\r 
\eea
where $Y_I$ is an arbitrary covariant function, and where  we also  defined the following tensors
\bea
U_I^J&=&\delta_I^J + \d_I B^J+ \half P^A_I (T_A\lambda)^J
\nl
 \gamma^I_{JK}&=&(U^{-1})^I_L\l \d_{(J}\gamma_{K)}^L  + P_{(J}^A(T_A)^L_{K)}\r~.
\eea
Notice that in the specific example of eq.~(\ref{eq_var_nabla_lambda}) the Weyl operator acts on $G_F$ singlets. Therefore the result is automatically dependent only on the invariant coefficient functions $B$ and $P$. In the case of the Weyl variation of tensors of $G_F$ there would appear an additional $G_F$ rotation with Lie parameter $S^A$.
In the course of our study we shall however mostly encounter the action on $G_F$ singlets.

\subsubsection{Source reparametrization and the form of $\Delta_\sigma$}
\label{sec_geometry}

The choice of parametrization of the sources is of course subject to some freedom. A change of parametrization leads to a change in the definition
of the renormalized composite operators and in the form of the Weyl operator $\Delta_\sigma$. Compatibly with dimensionality, one can consider the reparametrization
\bea
\label{reparsources}
{\lambda^I}'&=&\lambda^I +f^I\nl
{A^A_\mu}'&=&A^A_\mu+ f ^A_I\n_\mu\lambda^I \nl
{m^a}'&=&m^a +f^a_b m^b + \frac 1 6  f^a R + f_I^a \n^2\lambda^I +\half f^a_{IJ}\n_\mu\lambda^I \n^\mu\lambda^J~.
\eea
Provided the various coefficients $f_I, \,f^A_I,\,\dots$ respect $G_F$ covariance, the new parameters ${\lambda^I}',{A^A_\mu}',{m^a}'$ transform as the corresponding original ones under 
$G_F$. The effective action  changes form but its value is unaffected:
\bea
\WW' [g,\lambda',A',m']\equiv \WW [g,\lambda,A,m]\, .
\eea
The form of the Weyl operator in the new coordinates is straightforwardly derived by applying the chain rule. One finds the following relation for the coefficients in the new coordinate system:
\bea
\label{eq_scheme_dependence}
{\beta^I}'&=& \beta^I+\beta^J\d_J f^I\nl
{\rho_I^A}'&=&  \rho_I^A + \LL[ f_I^A] \nl
 {S^A}'&=&S^A - B^I f_I^A\nl
{ C^a}'&=& C^a- \frac 1 6  \LL[ f^a] \nl
{ D_I^a}'&=& D_I^a - \LL[ f_I^a] \nl
{ E_{IJ}^a}'&=&
E_{IJ}^a- \LL[ f^a_{IJ} ]- 2U^K_L \gamma^L_{IJ}f^a_K  
\nl
{ \theta_{I}^a}'&=& \theta_{I}^a +  B^J f^a_{JI} +2 U^J_If_J^a \nl
{ \eta^a}'&=&  \eta^a + f^a -B^I f_I^a
\eea
where we used the Lie derivative and the matrix $U_I^J$ introduced in the previous section.%

The most important remark concerning the above equation is that by  a suitable choice of $f^a$ and $f^a_I$, the tensor coefficients $\eta^a$ and $\theta^a_I$ can both be set to zero. As suggested by eq.~(\ref{TB}), and as further clarified  in section \ref{sec_improvement_and_SFT}, the choice $\eta^a=0$ corresponds to an ``improved" energy momentum tensor. 

As we said, the change of coordinates corresponds to a redefinition of the renormalized operators.
It is possible, however to find linear combinations of  operators that are invariant under the change of basis.
Consider, for example, the change of coordinates $m^a\to {m^a}'=m^a + f_I^a \n^2\lambda^I $. 
Focussing on the scalars for simplicity, the operators in the new basis are related to the original ones via
\bea
\label{oshift}
\left[\OO_I\right]
&=&\left[\OO_I\right]' + f_I^a \n^2 \left[\OO_a\right]'\nl
\left[\OO_a\right]&=&\left[\OO_a\right]'\eea
Combining this with eq. \eqref{eq_scheme_dependence} we find that the operator
\bea
\label{eq_scheme_independent_OI}
[\wt \OO_I]&=& \left[\OO_I\right] + \half (U^{-1})_I^J {\theta_J^a} \n^2 \left[\OO_a\right]
\eea
is scheme independent. This definition will be useful in section \ref{sec_LPR}.

\subsubsection{Consistency conditions}
\label{consistency_conditions}

The abelian nature of the Weyl symmetry imposes constraints on the form of the generator $\Delta_\sigma$. The vanishing of the commutator \bea
\label{eq_operator_consistency0}
\left [\Delta_{\sigma_2},\Delta_{\sigma_1}\right] = 0
\eea
leads to a set of equations relating the different coefficients appearing in (\ref{deltasource}):
\bea
\label{eq_consistency_condition}
B^I P_I^A&=&0\nl
B^I D_I^a ~ &=&~~\LL[\eta^a] + 6 C^a 
 \nl
B^J E_{JI}^a &=&- \LL [\theta_I^a]
-2 U^J_I D_J^a 
\eea
Notice that these consistency conditions  are independent of the choice of gauge discussed in section \ref{sec_ambiguity}.
Alternatively, as shown in appendix \ref{app_CS_derivation}, these conditions  can  be derived by directly computing the coefficients of $\Delta_\sigma$ from a dimensionally regulated action. According to that derivation the abelian nature of Weyl invariance, as realized on the bare sources in eq.~(\ref{baretransf}), is just an explicit fact, which need not be imposed.

One can easily check that the consistency condition $P^A_IB^I=0$ implies  $\LL[B^I]=0$. Together with $\LL[(T_A\lambda)^I]=0$ we thus have
\bea
B^I \LL [ Y_{IJ\ldots}]&=& \LL [ B^I Y_{IJ\ldots}] \nl
(T_B\lambda)^I  \LL [Y_{AI\ldots}]&=& \LL [ (T_B\lambda)^I Y_{AI\ldots}] 
\eea

What  role is played by eq.~(\ref{eq_consistency_condition})? For instance, at a point where $B=0$, the second equation ensures that, once the choice $\eta^a=0$ is made, $C^a$ must also vanish. Eq.~(\ref{RGflowoperators}) then implies that if $T$ is improved so as  to vanish at a given  RG scale  then it  automatically vanishes at all scales. 
 The first equation, as we shall see in section 3, ensures the absence of currents in the short distance singularities of correlators with multiple insertions of $T$. This  significantly simplifies the derivations of  constraints on the structure of  RG flows.

\subsubsection{Dimension 2 covariant functions}
\label{sec_dim_2_covariant}

In general, the Weyl transformation of dimensionful functions of the sources contains derivatives of $\sigma$ (see eq. \eqref{eq_var_nabla_lambda}). However, it is possible to find linear combinations of dimension 2 functions which transform ``covariantly" under this symmetry:
 \bea
 \label{eq_Pi}
 \Pi^{IJ}&=&\n_\mu \lambda^I\n^\mu \lambda^J -B^{(I} \Lambda^{J)}\nl
 \Pi^a&=& m^a -\eta^a \frac R 6 - \half  \theta_I^a \Lambda^I
 \eea
   where we defined the function
 \bea
 \Lambda^I&=&\l U^{-1}\r ^I_J\l  \n^2\lambda^J +\frac 1 6  B^J R \r ~.
 \eea
The variations of $\Pi^{IJ}$ and $\Pi^a$ contain no derivatives of $\sigma$. In the ``gauge" $S^A=0$ they are
\bea
\label{eq_pi_transformation}
\Delta_\sigma \Pi^{IJ}&=&\sigma \l 2\Pi^{IJ} - \gamma^I_K\Pi^{KJ} - \gamma^J_K\Pi^{IK}  + \gamma_{KL}^{IJ} \Pi^{KL}\r \nl
\Delta_\sigma \Pi^{a}&=&\sigma \l 2\Pi^{a} - \gamma^a_b\Pi^{b}  + \gamma_{IJ}^{a} \Pi^{IJ}\r 
\eea
 where   we defined the  tensors
 \bea
\gamma^{IJ}_{KL}&=&B^{(I} \gamma^{J)}_{KL}\nl
\gamma_{IJ}^{~a}&=&\half \l E_{IJ}^a + \theta^a_K\gamma^K_{IJ}\r ~. 
\eea
In computing the transformation property of $\Pi^a$ we imposed the consistency conditions \eqref{eq_consistency_condition}.
$\Pi^{IJ}$ and $\Pi^a$ will play an important role in the rest of the paper.

\subsubsection{Limiting cases}
\label{sec_improvement_and_SFT}

It is interesting to consider various limiting `fixed points'. 
Focusing on $T$ in eq.~(\ref{eq_T_beta_S_t}), we can basically consider three cases:
\begin{enumerate}
\item
When both $\eta^a$ and $B^I\equiv \beta^I-S^A(T_A\lambda)^I$ are zero the operator $T$ vanishes, 
corresponding to  a conformal fixed point.
Notice that conformality is signaled by the vanishing of $B^I$ and not of any other choice of $\beta$-function. Conformal theories with non-vanishing $\beta$-functions were discovered in \cite{Fortin:2012hn}. 

It is interesting to consider the conformal transformations in eq.~\eqref{eq_K_transformations} in this limit. Choosing a parametrization where $\theta_I^a=0$, the consistency conditions eq.~\eqref{eq_consistency_condition}  imply $D_I^a=C^a=0$, so that all
entries in eq. \eqref{eq_K_transformations} vanish,  apart from one.  In particular one finds $K^\mu {\cal O}^a=K^\mu J_\nu^A=0$, consistent with these operators being primaries,
but  also $K^\mu {\cal O}^I = -P_I^A J^{\mu}_A$, indicating that some of the $ {\cal O}^I$ are descendants of the currents. This result  is indeed expected because of eq.~\eqref{eq_naive_WI}.
In appendix \ref{app_p_at_CFT} we study this in detail showing  there exists an operator  basis where   each broken symmetry current is associated to its a unique scalar descendant.
In this basis all the remaining scalar operators are annihilated by the generator of special conformal transformations and all the remaining  currents are conserved and have vanishing anomalous dimension.

\item The case $B^I=0$  and $\eta^a\not=0$ corresponds to a fixed point whose energy-momentum tensor is not improved
\bea
T(x)&=& - \eta^a  \n^2  \left[\OO_a\right]~.
\eea
This possibility is relevant  when considering a QFT  flowing to different CFTs in the  UV and in the IR. Adjusting the coupling to the background metric such that the energy momentum tensor is improved at one asymptotic does not imply improvement at the other.

\item Another type of conceivable fixed point is an SFT, corresponding to the existence of a scheme where $\beta^I=0$ but $S_A\not =0$ so that $B^I\not = 0$. As noticed below \eqref{Kgenerator}, such point in coupling space is invariant under dilation but not under conformal transformations.
In this case \eqref{eq_T_beta_S_t}  becomes
\bea
 \label{eq_T_d_virial}
 T&=&-\n_\mu \left[V^\mu\right]
 \eea
 where $V^\mu=S^A J_A^\mu+\eta^a \n^\mu\OO_a $  is referred to as the virial current.  By  eq.~\eqref{TB}, since $B^I \not=0$ and since ${\cal O}_I$ and ${\cal O}_a$ are independent operators, we have also that $T =-\n_\mu V^\mu\not=0$, with no possibility of improvement to make $T=0$. The fact that $T$ vanishes only up to a total derivative is another way to see that the theory is endowed with  global scale invariance, but not with conformal invariance (local scale invariance).  Perturbative unitary SFTs are ruled out by the argument in ref. \cite{Luty:2012ww}, which we shall revisit in section \ref{sec_LPR}.
 
 Notice that in the case of an SFT, one can consistently consider a reduced set of sources by freezing $\lambda^I=\lambda^I_*= const$ such that $\beta^I=0$ and by reducing $A_\mu^A$ to a one dimensional subspace: $A_\mu^A\equiv S^A C_\mu$. One can then easily check that the Weyl transformation of $A_\mu^A$ in eq.~(\ref{deltasource}), simply reduces to $\delta_\sigma C_\mu = \n_\mu\sigma$. The relation $B^IP_I^A=0$ is essential to obtain this result.
 The source $C_\mu$ so defined thus corresponds to the virial current gauge field of ref. \cite{Luty:2012ww}. Notice also that the inhomogeneous terms in $\delta_\sigma m^a$, at $\eta^a=\theta^a_I=0$, package into a term proportional to $\tilde R\equiv R+6\n^\mu C_\mu-6C^\mu C_\mu$. Similarly the quantities $\Pi^{IJ}$ reduce to constant coefficients times $\tilde R$. The quantity $\tilde R$ on the reduced set of sources $g_{\mu\nu}, C_\mu$ satisfies  $\delta_\sigma \tilde R=2\sigma \tilde R$ and  plays an important role in the structure of the anomaly in a SFT, as we shall comment later.
 
 \end{enumerate}

\subsection{The structure of the Weyl anomaly}
\label{sec_anomaly}

We will now discuss the structure of the anomaly appearing in the local CS equation
\bea
\Dw_\sigma \WW [g,\lambda,m,A]&=& \int d^4x  \AA_\sigma
\eea
First, let us review the anomaly at an improved conformal fixed point ($B^I=\eta^a=0$). This case corresponds to  freezing all the sources apart from the metric ($\lambda^I=\lambda^I_*=const$, such that $B^I(\lambda_*)=0$ and $A^A_\mu=m^a=0$). The Weyl generator $\Delta_\sigma$ thus reduces to the metric variation $\Delta_\sigma^g$. The anomaly $\AA_\sigma$ is a linear combination of all the dimension 4 scalars that  can be constructed from the metric and its derivatives \cite{Capper:1974ic,Deser:1976yx}
\bea
\label{eq_Weyl_anomaly}
\frac {1}{\sqrt {- g}}\AA_\sigma &=&\sigma \l a{E_4}   - b R^2 -cW^2 \r -\n^2\sigma   d R ~.
\eea
where $R$ is the scalar curvature, $W^2$ is the Weyl tensor squared, and $E_4$ is the  4-dimensional Euler density.

The anomaly is constrained by  a Wess-Zumino integrability condition \cite{Wess:1971yu}:
since the Weyl symmetry is abelian, one must have
\bea
\Delta^g_{\sigma_2} \l \int  dx_1 \AA_{\sigma_1}\r-\Delta^g_{\sigma_1}\l \int  dx_2\AA_{\sigma_2 }\r&=&\left [\Delta^g_{\sigma_2},\Delta^g_{\sigma_1}\right]\WW
=0~.
\label{WZ}
\eea
This condition is satisfied by all terms in eq.~(\ref{eq_Weyl_anomaly}) apart from $R^2$. At a CFT  fixed point, the anomaly coefficient $b$ must therefore vanish.

Deser and Schwimmer classified the conformal anomalies into three types \cite{Deser:1993yx}: 
\begin{itemize}
\item
Contributions that equal the  variation of a local functional. 
Such contributions can be eliminated by adding to the action a suitable local functional. They must, therefore, not be considered as genuine anomalies. 
In the present case, $\n^2\sigma R$ corresponds to such a removable term, as it equals  the Weyl variation of  ${\sqrt g }R^2$.
\item
Type ``A": Anomalies that vanish when integrated over space-time with a constant $\sigma$. An equivalent characterization of these anomalies is that they do not contribute to
\beq
\mu \frac{d}{d\mu} {\cal W}\equiv \Delta^{RG} {\cal W}\,.
\label{RGW}
\eeq
Therefore  type ``A" anomalies are not associated with additional (logarithmic) UV divergences arising  in the presence of space-time dependent sources.
The Euler density anomaly is such an anomaly because its integral vanishes on topologically trivial spaces, such as Minkowski space. In practice this is because $\sqrt g E_4$
can be locally written as a total derivative (of a non covariant quantity).
\item
Type ``B":
Anomalies that do not vanish when integrated over space-time. Equivalently, by the previous argument involving $\Delta^{RG}$, these anomalies are associated with an explicit $\ln \mu$ dependence in the effective action. In 4D CFTs the corresponding anomaly is $W^2$. An example of the associated $\ln \mu$ dependence is given by the two point function of $T_{\mu\nu}$ which in Fourier space reads
\beq
\langle T_{\mu\nu}T_{\rho\sigma}\rangle =c \Pi_{\mu\nu\rho\sigma}^{(2)}p^4 \ln p^2/\mu^2\, ,
\eeq
where $\Pi_{\mu\nu\rho\sigma}^{(2)}$ is the projector on transverse traceless 2-index tensors.
\end{itemize}
Strictly speaking, also the $E_4$ can  give rise to a $\ln\mu$ dependence, but only when the CFT is embedded in a space with non trivial topology, like for instance the sphere $S_4$. In any case, the logarithmic divergences associated with $E_4$ do not affect local quantities, such as correlators.

Let us now consider the anomaly  in the presence of all the external sources,  and see what becomes of the properties we just discussed.
Up to terms involving $\epsilon^{\mu\nu\rho\sigma}$, the most general form, first  given in \cite{Osborn:1991gm}, is
\bea
\label{eq_osborns_anomaly}
\frac {1}{\sqrt {- g} }\AA_\sigma
&=&\sigma \l \beta_a W^2 + \beta_b E_4+\frac 1 9 \beta_c R^2\r 
- \n^2\sigma \l  \frac 1 3   dR\r  \nl
&&+
\sigma \Big(
 \frac 1 3 \chi_I^e \n_\mu \lambda^I \n^\mu R  
+\frac 1 6 \chi_{IJ}^f \n_\mu \lambda^I \n^\mu\lambda^J R  
+\frac 1 2 \chi_{IJ}^g G^{\mu\nu}\n_\mu \lambda^I \n_\nu\lambda^J  
\nl
&&~~~~~
+\frac 1 2 \chi_{IJ}^a\n^2 \lambda^I\n^2 \lambda^J 
+\frac 1 2 \chi_{IJK}^b\n_\mu \lambda^I\n^\mu \lambda^J \n^2 \lambda^K
 +\frac 1 4 \chi_{IJKL}^c\n_\mu \lambda^I\n^\mu\lambda^J \n_\nu\lambda^K \n^\nu \lambda^L
\nl
&&
+\n^\mu \sigma \Big(
G_{\mu\nu}w_I\n^\nu \lambda^I
+ \frac 1 3 R Y_I \n_\mu \lambda^I
+\wt S_{IJ}\n_\mu\lambda^I (U^{-1})^J_K\n^2 \lambda^K
+\half T_{IJK} \n_\nu\lambda^I\n^\nu\lambda^J \n_\mu\lambda^K \Big)\nl
&&-\n^2\sigma  \l U_I \n^2\lambda^I + \half V_{IJ} \n_\nu\lambda^I \n^\nu \lambda^J\r\nl
&&+\sigma \Big(\half p_{ab}\hat m^a \hat m^b+\hat m^a \l \frac 1 3 q_aR + r_{aI}\n^2 \lambda^I 
+\half s_{aIJ}\n_\mu\lambda^I \n^\mu \lambda^J\r\Big)\nl
&&+
\n_\mu \sigma  \l \hat m^a j_{aI}\n^\mu \lambda^I  \r 
-  \n^2\sigma \l   \hat m^a k_a\r \nl
&&+\sigma \l \frac 1 4 \kappa_{AB} F_{\mu\nu}^A F^{B\mu\nu} + \half \zeta_{AIJ}F_{\mu\nu}^A \n^\mu\lambda^I \n^\nu\lambda^J\r 
+ \n^\mu \sigma \l \eta_{AI}  F^A_{\mu\nu} \n^\nu \lambda^I\r 
\eea
where $G_{\mu\nu}$ is the Einstein tensor, $F_{\mu\nu}^A$ is the field strength associated with the background field $A_\mu^A$ and $\hat m^a=m^a-\frac{1}{6}\eta^aR$.
As in the CFT limit, ${\cal A}_\sigma$ is redundant, in that  it is only defined modulo  the variation of a local functional $F$ of the sources: ${\cal A}_\sigma\sim {\cal A}_\sigma+\Delta_\sigma F$. This redundancy  corresponds to the freedom in choosing a renormalization procedure.
 At the same time  ${\cal A}_\sigma$ is
 subject to the Wess-Zumino  consistency condition, now given by the analogue of eq.~(\ref{WZ}) with $\Delta_\sigma$ instead of $\Delta^g_\sigma$.
This condition  translates \cite{Osborn:1991gm} into $\sim 10$ differential equations involving the $25$ tensor coefficients appearing in $A_\sigma$. 

A new result, which we present here, is a reformulation of the anomaly, in which most of the consistency equations  are explicitly solved, leaving only three non-trivial constraints. One of these is  the equation discovered in \cite{Jack:1990eb,Osborn:1991gm} and describing the flow of the coefficent $a$. The other two equations involve instead the anomaly coefficients associated with the flavor gauge fields. One key observation in our analysis is that, by eliminating  a suitable set of  scheme dependent terms, most of the consistency equations
become  algebraic. They can thus be readily solved and   substituted back into the anomaly. The consistency equations in this suitable scheme choice appear in appendix \ref{app_osborn_anomaly}.
 
 According to our analysis the general anomaly in eq.~(\ref{eq_osborns_anomaly}) can be written as a sum of five terms which we indicate using an analogy with the Weyl anomaly of a CFT (eq. (\ref{eq_Weyl_anomaly})):
\bea
\label{eq_Weyl_anomaly_Off_criticality}
\AA_\sigma&=&\AA_\sigma^{\n^2 R} + \AA_\sigma^{R^2} + \AA_\sigma^{W^2}+\AA_\sigma^{E_4}+\AA_\sigma^{F^2}~.
\eea
The different parts of the anomaly are:
\begin{enumerate}
\item {Generalized $\n^2 R$ anomaly}

The generalized $\n^2R$ anomaly represents the terms that can be written as $\Delta_\sigma F$ and can thus be eliminated by a choice of scheme.
By a proper choice of local terms, that is specified in the appendix,  the coefficients
$d, U_I, V_{IJ}, \wt S_{(IJ)}, T_{IJK}, k_a, j_{aI}$  can be set to zero.

\item {Generalized $R^2$ anomaly}

The terms associated with  $\beta_c, Y_I, \chi_I^e, \chi_{IJ}^f,\chi_{IJ}^a,\chi_{IJK}^b, q_a, r_{aI}$ can be rewritten using the  consistency equations in  the following compact form: 

\bea
\label{eq_generalized_R2}
\frac {1}{\sqrt {- g} }\AA_\sigma^{R^2}&=&\sigma\l  \half b_{ab} \Pi^a \Pi^b
+ \half  b_{aIJ} \Pi^a \Pi^{IJ}+\frac 1 4 b_{IJKL} \Pi^{IJ}\Pi^{KL} \r
\eea
This part of the anomaly is simply the most general bilinear scalar constructed from the covariant objects $\Pi^{IJ}$ and $\Pi^a$ which were defined in \eqref{eq_Pi}.
 Since the variation of the $\Pi$'s does not contain derivatives of $\sigma$, the above term is  manifestly consistent.

 We refer to this anomaly as the generalized $R^2$ anomaly because in the limit where $\n\lambda=m=0$ the only term remaining from this anomaly is proportional to $R^2$. 
  The definitions of the coefficients appearing here, in terms of the original parameterization of the anomaly, are given in the appendix.

\item {Generalized $W^2$ anomaly}
\bea
\frac {1}{\sqrt {- g} }  \AA_\sigma^{W^2}&=&-\sigma c ~W^2
\eea
The form of the $W^2$ anomaly is unchanged off criticality. The only difference is that the $c$ coefficient is replaced by a function of the sources $\lambda^I$, but the anomaly remains manifestly consistent.

\item {Generalized $E_4$ anomaly}

As in the case of the $W^2$ anomaly, away from the fixed point, the coefficient of the $E_4$ anomaly is  a function of the $\lambda$'s, and is thus space-time dependent. However, since the Weyl variation of $E_4$ contains two derivatives of $\sigma$, the consistency condition involves (after integration by parts) terms proportional to  $\n_\mu a$, which  are not present at the fixed point where $a(\lambda)$ is  a numerical constant. The result is that the $E_4$ anomaly is no longer  automatically consistent away from criticality: additional terms must exist in order to restore consistency. We find that a consistent anomaly containing $E_4$ must have the following structure:
\bea
\label{eq_chi_g_w_anomaly}
\frac {1}{\sqrt {- g} } \AA_\sigma^{E_4} &=&
\sigma  \l a E_4+ \chi_{IJ}^g \l \half \Gamma_{\mu\nu}    \n^\mu\lambda^I \n^\nu \lambda^J- \frac 1 4  U^I_K \Lambda^K \Lambda ^J\r
+ \half \overline \chi_{IJK} ^g   \Omega^{IJK} 
\r 
\nl
&& + \n^\mu \sigma  \l w_I G_{\mu\nu}\n^\nu \lambda^I\r   - \half \d_{[J}w_{I]}\Xi^{IJ}_\sigma
\eea

 where $\chi_{IJ}^g$ and $w_I$ are functions of $\lambda$, introduced in eq.~\eqref{eq_osborns_anomaly}, and where we used the notations defined in sec. \ref{sec_dim_2_covariant} plus the definitions
\bea
\Gamma_{\mu\nu}&=&G_{\mu\nu} + \frac R 6  g_{\mu\nu}\nl
\Omega^{IJK}&=&\l \Pi^{IJ} +  \half B^{(I} \Lambda ^{J)}\r \Lambda ^K\nl
\Xi^{IJ}_\sigma&=&\Lambda^I \l 2\n_\mu\sigma \n^\mu\lambda^J - \sigma\gamma^J_{KL}\Pi^{KL} \r~. 
\eea
and 
\bea
\overline \chi^g_{IJK} &=& - \d_{(J}\chi^g_{KI)} +  \half   \d_K\chi_{IJ}^g~.
\eea
Notice that, even though it involves  several terms, this  anomaly  is  described by  just three tensor functions $a, w_I, \chi_{IJ}^g$.  Moreover, Wess-Zumino  consistency implies the following constraint
\bea
\label{eq_w_I_consistency} 
\LL[ w_I]&=&-8\d_Ia+\chi_{IJ}^gB^J
\eea

\item {Generalized $F^2$ anomaly}

The generalized $F^2$ anomaly depends on three coefficients, $\kappa_{AB},\zeta_{AIJ}$ and $\eta_{AI}$, and takes the  form
 \bea
 \label{eq_F2_anomaly}
\frac {1}{\sqrt {- g} } \sigma \AA_\sigma^{F^2}&=&
\sigma \l \frac 1 4 \kappa_{AB} F_{\mu\nu}^A F^{B\mu\nu} 
+ \half \zeta_{AIJ} F_{\mu\nu}^A \n^\mu\lambda^I \n^\nu\lambda^J+ \l  \half  P^A_I \zeta_{AJK} + \eta_{AI} \d_{[J} P^A_{K]} \r \Omega^{IJK} \r \nl
&&+ \n^\mu \sigma \l \eta_{AI}  F^A_{\mu\nu} \n^\nu \lambda^I\r 
-\half \eta_{A[I} P^A_{J]} \Xi^{IJ}_\sigma
\eea
The three coefficients appearing in this anomaly are related to one another and to the coefficients of the generalized $E_4$ anomaly via 2 consistency conditions\footnote{Indeed the $E_4$ anomaly  is not fully consistent on its own in the presence of a non-vanishing field strength background $F^A_{\mu\nu}$. Terms involving the field strength in the Weyl variation of the $E_4$ anomaly go along with similar terms from the $F^2$ anomaly, and thus appear in the $F^2$ consistency condition in eq.~(\ref{eq_F2_CC}).}
\bea
\label{eq_F2_CC}
\LL[ \eta_{AI}]&=&\kappa_{AB}P^B_I+\zeta_{AIJ}B^J- \chi^g_{IJ} (T_A\lambda)^J\nl
0&=&\eta_{AI}B^I+ w_I (T_A\lambda)^I
\eea
\end{enumerate}

In the end we find that the anomaly can be described by  10 physical scheme independent tensorial coefficients, constrained by the 3 consistency conditions in eqs.(\ref{eq_w_I_consistency},\ref{eq_F2_CC}). Note however that the second constraint in \eqref{eq_F2_CC} is not fully independent from the other two. Indeed, the vanishing of the Lie derivative of this constraint is automatic once the other two constraints are enforced.

\subsubsection{Comments on the $R^2$ anomaly}
\label{sec_R2}

Some comment on the ${\cal A}^{R^2}_\sigma$ anomaly are in order, as it represents a novelty compared to the well known CFT limit. We will show that it is associated with logarithmic divergences  in  CFTs that can be ``unimproved" when scalar operators of dimension exactly equal to two are present. We will also show that the components associated with operators with non-zero anomalous dimensions can be eliminated by a choice of scheme.

The coefficients $b_{ab}$, $b_{aIJ}$ and $b_{IJKL}$ are associated with the short distance singularities in respectively $\langle {\cal O}_a{\cal O}_b\rangle$, $\langle {\cal O}_a{\cal O}_I{\cal O}_J\rangle$ and $\langle {\cal O}_I{\cal O}_J{\cal O}_K{\cal O}_L\rangle$. This is easily seen by considering the action of the RG flow operator $\Delta^{RG}$.  For instance concerning $G_{ab}(x,y)=i\langle {\cal O}_a(x){\cal O}_b(y)\rangle$ one has
\bea
\label{abRG}
\mu\frac{d}{d\mu}G_{ab}(x,y)&=&\Delta^{RG}\frac{\delta}{\delta m_a(x)}\frac{\delta}{\delta m_b(y)}{\cal W}\\
&=&[\Delta^{RG},\frac{\delta}{\delta m_a(x)}\frac{\delta}{\delta m_b(y)}]{\cal W}+\frac{\delta}{\delta m_a(x)}\frac{\delta}{\delta m_b(y)}{\cal A}_{-1}\\
&=&-\gamma_{a}^cG_{cb}(x,y)- \gamma_{b}^cG_{ac}(x,y)-b_{ab}\delta^4(x-y)\,.
\eea
The implications of this equation are more easily worked out  in momentum space. For instance, at the original unperturbed CFT fixed point where $\gamma_a^b,B^I=0$ and  $b_{ab}\equiv b_{ab}^{(0)}=const$, we have
\beq
\mu \frac{d}{d \mu} G_{ab}(p^2)=-b_{ab}^{(0)} \qquad\longrightarrow \qquad G_{ab}=-\frac{1}{2} b_{ab}^{(0)} \ln (-\mu^2/p^2)
\label{b_ab_log}
\eeq
where the cut of the logarithm is chosen along the positive real axis $p^2\geq 0$. Now for the imaginary part at $p^2 = |p^2|+i\epsilon$  we find $2\,{\mathrm {Im}} \,G_{ab}= -\pi b_{ab}^{(0)}\theta(p^2)$ so that by unitarity we conclude that   $b_{ab}^{(0)}$ must be negative definite. Considering the expression for $\Pi^a$ in eq.~(\ref{eq_Pi}) at the original fixed point $\lambda^I=0$,  the anomaly associated with $b_{ab}^{(0)}$ reduces to
\beq
b_{ab}^{(0)} \left (m^a-\frac{\eta^a}{6} R\right ) \left (m^b-\frac{\eta^b}{6} R\right )\, .
\label{b_ab_CFT}
\eeq
By eq.~\eqref{b_ab_log} this result is readily intepreted as due  to a deformation of the CFT by the  coupling $(m^a-\eta^a R/6){\cal O}_a$. This  is also consistent with the interpretation of $\eta^a$ as a parameter describing the ``unimprovement" of the CFT. We stress, although it is obvious, that compared to the standard CFT anomaly in eq.~\eqref{eq_Weyl_anomaly}, where $R^2$ is inconsistent, eq.~\eqref{b_ab_CFT} is made consistent by the Weyl tranformation of an extra source, $m^a$.
A related discussion of this issue is found in sect. 2.3 in ref. \cite{Luty:2012ww}. 

Notice that the coefficients  $b_{ab}$, $b_{aIJ}$ and $b_{IJKL}$ can be modified by the addition of  local counterterms of the same form:
\bea
\delta\WW&=&\int d^4x \sqrt {-g} \l \half c_{ab} \Pi^a \Pi^b
+ \half  c_{aIJ} \Pi^a \Pi^{IJ}+\frac 1 4  c_{IJKL} \Pi^{IJ}\Pi^{KL}\r \nl 
\delta b_{ab}&=&
  -\LL\left[   c_{ab}\right] 
 \nl
\delta b_{aIJ}&=&
 -\LL\left[ c_{aIJ}\right]  +\gamma^{KL}_{IJ} c_{aKL}
+ 2\gamma_{IJ}^b c_{ab} \nl 
\delta b_{IJKL}&=&
 -\LL\left[  c_{IJKL}\right]
 +  \gamma_{IJ}^{MN} c_{MNKL}  
+  \gamma_{KL}^{MN} c_{IJMN}+  \gamma_{KL}^a c_{aIJ} +\gamma_{IJ}^ac_{aKL}~.
\eea
In particular, at a CFT fixed point $\delta b_{ab}=\gamma^c_ac_{cb} + \gamma^c_bc_{ac}$, so that all the entries in $b_{ab}$ can be eliminated apart from those associated with operators of dimension exactly equal to 2. This makes sense because only for those entries  does $ G_{ab}(p^2)$ involve a logarithm, corresponding to an ineliminable $\ln \mu$ dependence in ${\cal W}$. The same remark applies to $b_{aIJ}$ and $b_{IJKL}$: around a CFT fixed point the only genuine anomalies, the ones that cannot be removed by local counterterms, correspond to 3- and 4-point functions of fields, such that the sum of their anomalous dimensions vanishes. 

It is also interesting to consider what would become of these anomalies in the limit of an SFT. Limiting  the set of sources to just $g^{\mu\nu}$ and the virial gauge field $A^A_\mu = S^A C_\mu$, and improving the theory by the choice $\eta^a=0$, the anomaly reduces to a term proportional to ${\tilde R}^2$ (see sec. \ref{sec_improvement_and_SFT}). This is the SFT anomaly discussed in ref. \cite{Luty:2012ww}. As this anomaly coefficient controls the $J=0$ component of the energy momentum 2-point function, one easily deduces that the coefficient must be positive in a unitary theory.

\subsection{Weyl consistency conditions and gradient flows}
\label{sec_gradient_flow}
If one considers the quantity \cite{Jack:1990eb,Osborn:1991gm}
\beq
\tilde a =a +\frac{1}{8}w_I B^I
\eeq
then eq.~(\ref{eq_w_I_consistency}) together with the second constraint in eq.~(\ref{eq_F2_CC}) implies the famous gradient flow equation
\beq
8\partial_I \tilde a=\left (\chi_{IJ}^g+\partial_I w_J-\partial_J w_I+P^A_I\eta_{AJ}\right ) B^J\,.
\label{our gradient flow}
\eeq
 The gradient flow equation is one  major result in the work of Jack and Osborn\cite{Jack:1990eb}. To our knowledge, however, in the general case involving global symmetries, it was not cast in the form of eq.~(\ref{our gradient flow}) until recently in \cite{Jack:2013sha}
 (see for instance section 3.6 of ref.\cite{Luty:2012ww}). Notice indeed that, in order to obtain eq.~(\ref{our gradient flow}),  eq.~(\ref{eq_F2_CC}) is  crucial, in that it implies that a seemingly spurious term
$P_I^A w_J (T_A\lambda)^J$ is indeed proportional to the $B^I$'s.  Eq.~(\ref{our gradient flow}) gives rather non-trivial relations among perturbative expansion coefficients of  the $\beta$-function and of the other quantities in the right hand side. Indeed, as pointed out in  \cite{Jack:1990eb}  and further demonstrated in \cite{Jack:2013sha}, there arise relations purely involving the $\beta$-functions of different couplings at different perturbative orders. For instance, in weakly coupled gauge theories with scalars, one can relate the leading contribution of the scalar quartic coupling to the gauge $\beta$-function, which comes at 3-loops, to the 1-loop $\beta$-function for the scalar quartic itself.

Another implication of eq.~(\ref{our gradient flow}) is that $\tilde a$ is stationary at a conformally invariant fixed point, where $B^I=0$. Notice that at a  CFT $\tilde a$ and $a$ have the same value, though  $a$ is in general not stationary. However, since  at a CFT $\partial_I a=-w_J\partial_I B^J/8$, we have that $a$  is still stationary with respect to marginal perturbations, that is perturbations associated with vanishing eigenvalues of $\partial_I B^J$. A corollary of this result is that $a$ must be constant on any manifold of fixed points. Moreover, since in a CFT $a$ is the coefficient of one of the three structures describing the 3-point function of $T_{\mu\nu}$ \cite{Osborn:1993cr}, our result implies the vanishing of the tensor structure corresponding to $a$ in
\beq
\int d^4x \langle  {\cal O}(x) T_{\mu\nu}(y)T_{\rho\sigma}(z)T_{\tau\chi}(w)\rangle\, .
\eeq
Although we have not studied that, this result should also be obtained by using the constraints imposed by conformal symmetry on the correlators.
A corresponding result applies in 2D CFTs for the the correlator $\int  d^2x \langle  {\cal O}(x) T_{\mu\nu}(y)T_{\rho\sigma}(z)\rangle$. Though in that case it trivially follows from the vanishing of correlators involving $n$ insertions of $T$ and one insertion of another primary, which is  a consequence of the Virasoro algebra.

However, the most interesting consequence of eq.~(\ref{our gradient flow}) is obtained by contracting it with $B^I$
\beq
8\mu\frac{d\tilde a}{d\mu}\equiv 8 B^I\partial_I \tilde a= \chi_{IJ}^g B^IB^J,
\label{mudmua}
\eeq
where the relation $B^I P_I^A=0$ was used. The relevance of this result lies in the positivity property of the matrix $\chi_{IJ}^g$, as for $\chi_{IJ}^g >0$ it implies  $\tilde a$ is a monotonically evolving function of the couplings. 
Moreover, in an SFT, one would have that $B^I\d_I=N^A(T_A\lambda)^I\d_I$ is just a $G_F$ rotation. Then 
the $G_F$ covariance of $\tilde a$ would  imply $\chi_{IJ}^g B^IB^J=0$. For a positive definite $\chi_{IJ}^g$ one would conclude that $B^I=0$, and that therefore the theory must be a CFT.

Indeed, as noted already in \cite{Jack:1990eb}, unitarity guarantees the positivity of $\chi_{IJ}^g$ in a neighborhood of the original CFT
where all $\beta$-functions and anomalous dimensions remain small. This proof is based on the following relation between $\chi_{IJ}^g$ and the anomaly coefficient $\chi_{IJ}^a$ (see eq. \eqref{eq_osborns_anomaly}):
\beq
\chi_{IJ}^g = -2\chi_{IJ}^a +O(B,\partial B, P)
\eeq
This relation can be derived from the Wess-Zumino consistency condition of the original anomaly. When $B,\partial B, P$ can be treated as perturbations, then  all anomalous dimensions are small and the positivity of $\chi_{IJ}^g$ coincides with negativity of $\chi_{IJ}^a$. We will now describe a proof for the negativity of this matrix in unitary theories. In section \ref{sec_LPR} we will present an alternative argument for the positivity of $\chi_{IJ}^g$ based on the dilaton scattering amplitude.  

The negativity of $\chi_{IJ}^a$ can be established as follows: by the same considerations used in the discussion around eq.~(\ref{abRG}) and by the use of eq.~(\ref{eq_osborns_anomaly}), we have that the two point function
$G_{IJ}\equiv i\langle {\cal O}_I(p){\cal O}_J(-p)\rangle$ satisfies the RG equation
 \beq
\mu \frac{d}{d \mu} G_{IJ}+\gamma_I^KG_{KJ}+\gamma_J^KG_{IK}=-(p^2)^2 \chi^a_{IJ} \label{IJRG}\,.
\eeq
Defining $t\equiv \frac{1}{2}\ln (-\mu^2/p^2)$, in such  a way that $t$ is real at euclidean momenta, the general solution of the above RG equation can be written in terms of the running couplings $\lambda^I(t)$ as
\beq
G_{IJ}=(p^2)^2\left [F_{IJ}(\lambda(0))+\int_0^t dt' (A(t') \chi^a(\lambda(t')) A^T(t'))_{IJ}\right ]
\label{solutionRG}
\eeq
where 
\beq
A(t)_I^J=\left ({\mathbf T} e^{\int _0^t dt' \gamma(\lambda(t')) }\right) _I^J
\eeq
and where $F_{IJ}$ is a scheme dependent integration constant. From the above equations, it follows that any RG scale, up to corrections controlled by the anomalous dimensions and $\beta$-functions,
\beq
{\mathrm {Im}}\, G_{IJ}=- \pi \chi_{IJ}^a\, .
\eeq
As long as those corrections can be neglected, unitarity implies $\chi_{IJ}^a<0$ and thus, by the previous discussion, $\chi^g_{IJ}$ must be positive.  Notice that this conclusion is not affected by changes of scheme generated  by the addition of local counterterms to the action. Indeed under these additions 
one has $\chi_{IJ}^a\to \chi_{IJ}^a+{\cal L} [c_{IJ}]$,  with $c_{IJ}$ a covariant function of the couplings: the change in $\chi_{IJ}^a$ is again controlled by anomalous dimensions
and $\beta$-functions, which are small under  our hypothesis. Let us stress again our conclusion: in a neighbourhood of the original fixed point (see fig. 1) where the $\beta$-function and the anomalous dimensions of ${\cal O}_I,{\cal O}_a, J_A^\mu$ can be treated as small perturbations, unitarity implies the positivity of $\chi_{IJ}^g$. We should also emphasize that this result does not rely on the perturbativity of $\lambda^I$. Indeed $\chi_{IJ}^g$ may differ significantly by its value at the fixed point, but under our assumptions of small $\beta$ and small anomalous dimensions, unitarity nails $\chi_{IJ}^g$ to be positive. Nonetheless, we understand that the generic situation is  one where the smallness of $\beta$ and of the anomalous dimensions is controlled by the size of the couplings $\lambda^I$ themselves.

Now, the integral of eq.~(\ref{mudmua}) 
\beq
\tilde a(\lambda(\mu_2))-\tilde a(\lambda(\mu_1))=\frac{1}{8}\int_{\mu_1}^{\mu_2} \chi_{IJ}^g(\lambda(\mu))B^I(\lambda(\mu))B^J(\lambda(\mu))\, d\ln \mu
\label{integralbound}
\eeq
gives a straightforward bound on the asymptotics of the RG flow. As long as the RG trajectory is  in the neighbourhood of the original fixed point, the left-hand side of eq.~(\ref{integralbound}) is finite, since  as   $\tilde a$ is a finite function of the renormalized couplings. Then, if the RG trajectory remains in this neighbourhood asymptotically,  $\ln \mu \to \pm \infty$, the positive  integrand at the right hand side must vanish in the corresponding asymptotics
\beq
\lim_{\ln\mu \to \pm\infty} \,\chi_{IJ}^g(\lambda(\mu))B^I(\lambda(\mu))B^J(\lambda(\mu))=0\, .
\label{Bsquared0}
\eeq
This can only happen if either $B^I\to 0$ or if $\chi_{IJ}^g$ asymptotes a matrix with null eingenvalues. In the latter case,
the operators corresponding to such eigenvalues would vanish in the limit where $\beta$-functions and anomalous dimensions are neglected: so they must vanish for real otherwise our hypothesis of negligible $\beta$-functions and anomalous dimensions is violated. We conclude that within our hypothesis, one must have  $B^I\to 0$ asymptotically for all non-null operators. The asymptotics must therefore be CFTs. A particular case satisfiying our hypothesis is that of Banks-Zaks type
theories: the only possible asymptotics in a neighbourhood of the original free field theory must as well be CFTs. 

\subsubsection{Gradient flow for the vector $\beta$-functions?}
Recently, it was conjectured by Nakayama \cite{Nakayama:2013ssa} that the vector $\beta$-function, defined as
$B_\mu^A \equiv  P^A_I\n_\mu\lambda ^I$ satisfies a form of a gradient formula
\bea
\label{eq_Nakayama}
B_\mu^A &\equiv &g_{\mu\nu} H^{AB}(\lambda) \frac{\delta}{\delta A_\nu ^B} f(A,\lambda)
\eea
with unknown functions $H^{AB}$ and $f$. 
By dimensional analysis and covariance $f$ must take the form $f(A,\lambda)=\half F_{IJ}\n_\mu\lambda^I \n^\mu\lambda^J$, and this equation can be rewritten in terms of the function $P_I^A$ as follows
\bea
\label{eq_conjecture_Nakayama}
P_I^A &=& H^{AB}F_{IJ}(T_B\lambda)^I
\eea
where $F_{IJ}$ is necessarily a symmetric tensor.

What can one say about this conjecture based on the Weyl consistency conditions?
 Eq. \eqref{eq_F2_CC}  can be recast in the following form:
\bea
P_I^A&=& 
(\kappa^{-1})^{AB} \l \chi_{IJ}^g+\partial_I w_J-\partial_J w_I+P^A_I\eta_{AJ}+P^A_J\eta_{AI}\r (T_B\lambda)^J \nl
 &&+(\kappa^{-1})^{AB}\l  \zeta_{BIJ}  + 2\d_{[I}\eta_{BJ]}\r B^J~.
\eea
Notice that by unitarity the matrix $\kappa_{AB}$ is invertible. Moreover, the tensor in the first line is equal to the tensor appearing in the gradient flow formula of the $\beta$ function \eqref{our gradient flow} (up to terms which are eliminated using $B^IP_I^A=0$).
We see that the conjecture \eqref{eq_conjecture_Nakayama} would imply non-trivial constraints on the $\beta$-function and anomaly coefficients, constraints for which we do not find evidence in the framework discussed here. 

In ref. \cite{Nakayama:2013ssa}, the validity of eq.~(\ref{eq_Nakayama}) was checked in CFTs based on an AdS dual.
As that result corresponds to the leading order in a $1/N$ expansion, we are tempted to guess it should not hold true when including higher orders.

\section{Correlation functions of $T$ off-criticality}
\label{sec_dilaton}

\subsection{The dilaton effective action}
\label{dilaton_action}

In this section we shall use the local Callan-Symanzik equation to write the correlators of $T$ in terms of the correlators of the other composite operators, plus local terms associated with the anomaly.
For this purpose we will introduce the dilaton field $\tau(x)$, and define the dilaton effective action $\Gamma[\bar g, \tau]$ 
as the quantum effective action $\WW$ evaluated in the background\footnote{We keep a non-trivial background metric in order  to allow in principle to control matrix elements of $T_{\mu\nu}$. But we shall eventually focus on the flat case $\bar g_{\mu\nu}= \eta_{\mu\nu}$.}
 \beq
{\cal J}_1(\bar g,\tau)\,\equiv\, (g^{\mu\nu}=e^{2\tau} \bar g^{\mu\nu} ,\,\lambda^I=\lambda^I(\mu)={\mathrm{const}}, \,A_\mu^A=0, \, m^a=0)\,.
 \label{eq_J_1}
\eeq 
This effective action can be written as an expansion in powers of $\tau$
\bea
\label{eq_diltaon_effective_action}
\Gamma [\br g,\tau]&=&\WW[\JJ_1]=\exp\left\{\Delta^g_\tau\right\}\WW[\JJ]\Big|_{\JJ=\JJ_0}=\sum_{n=0}^\infty 
\frac{1}{n!} \underbrace{\Delta_{\tau}^g \ldots \Delta_{\tau}^g}_n\WW[\JJ]\Big|_{\JJ=\JJ_0}
\eea
where we used the operator $\Delta^g_\tau$ defined in \eqref{eq_local_CS_definition}, and defined the background $\JJ_0$ as
\beq
{\cal J}_0(\bar g)\,\equiv\, (g^{\mu\nu}=\bar g^{\mu\nu} ,\,\lambda^I=\lambda^I(\mu)={\mathrm{const}}, \,A_\mu^A=0, \, m^a=0)\,.
 \label{eq_J_0}
\eeq 
Using the definition \eqref{nT}, we see that the coefficient of the $\tau(x_1)\dots\tau(x_n)$ term in $\Gamma [\br g,\tau]$, evaluated with a flat metric $\bar g^{\mu\nu}=\eta^{\mu\nu}$, 
corresponds to the n-point correlator for $T$
\bea
\Gamma[\eta,\tau]&=&\sum_{n=0}^\infty 
\frac{i^{n-1}}{n!}
\int d^4 x_n\ldots\int d^4x_1 ~\tau(x_n) \ldots \tau(x_1) \TO{ T (x_1) \ldots T (x_n) }~.
\eea

In order to write the correlators of $T$ in terms of those of  the other composite operators we need to consider the quantum action for the Weyl transformed sources
\beq
{\cal J}_2(\bar g,\tau)\equiv 
\exp\left\{-\Dw_\tau\right\} {\cal J}\Big|_{{\cal J}={\cal J}_1}= 
( \bar g^{\mu\nu} ,\,\lambda^I[\tau], \,A_\mu^A[\tau], \, m^a[\tau])\, ,
\eeq
for which the $\tau$ dependence is transferred to $\lambda^I,A_\mu^A,m^a$. We shall discuss below  the form of the  Weyl transformed sources $\lambda^I[\tau], \,A_\mu^A[\tau], \, m^a[\tau]$.
 The effective action for $\tau$ can then be conveniently written as the sum of two contributions
 \beq
\label{eq_dilaton_eft_exponent}
\Gamma [\bar g,\tau]= \Bigl \{\WW[{\cal J}_1]-\WW[{\cal J}_2]\Bigr \}+\WW[{\cal J}_2]\equiv\Gamma_{local}[\tau]+\Gamma_{non-local}[\tau]
\eeq
where the term in curly brackets $\equiv \Gamma_{local}$ is clearly local, as it corresponds to a finite Weyl variation of the action. The second term $\Gamma_{non-local}$ is
a functional where $\lambda^I[\tau], \,A_\mu^A[\tau], \, m^a[\tau]$ act as sources for respectively ${\cal O}_I, J_A^\mu, {\cal O}_a$. 
When focussing on an order by order expansion in $\tau$, it is also convenient to write eq.~\eqref{eq_dilaton_eft_exponent}  as
\bea
\Gamma [\bar g,\tau]&=&
\exp\{\Delta^g_\tau\}\l1 -\exp\left\{-\Dw_\tau\right\}\r\WW\Big|_{\JJ_0} 
+\exp\left\{\Delta^g_\tau\right\} \exp\left\{\Delta^\beta_\tau-\Delta^g_\tau\right\}\WW\Big|_{\JJ_0}
\nl
&=& 
\exp\left\{\Delta^g_\tau\right\} \l1 -\exp\left\{-\Dw_\tau\right\}\r\WW\Big|_{\JJ_0}
+\exp\left\{\Delta_\tau^\beta+\half \left[ \Delta^g_\tau, \Delta^\beta_\tau-\Delta^g_\tau\right]+\ldots\right\}\WW\Big|_{\JJ_0}\nl
&\equiv&\Gamma_{local}[\tau]+\Gamma_{non-local}[\tau]~.
\label{exponentials}
\eea
where in the second term we made use of $- \Delta _\tau=\Delta^\beta_\tau-\Delta^g_\tau$.
In principle, the dots in the second line can be completed using the Baker-Campbell-Hausdorff (BCH) formula.
Again, the first term is manifestly local because all the terms in it involve at least one power of $\Delta_\tau$ acting on $\WW$, which 
gives the anomaly $\AA$.
The second expression is a series of terms involving derivatives of $\WW$  with respect to the sources, that is a
series of correlation functions of composite operators. 
Notice that in the absence of dimension 2 operators, all the commutators in the BCH formula vanish, and the computation simplifies significantly. 

In principle, the effective action can be obtained by working out the exponentials in eq.~\eqref{exponentials} order by order in $\tau$. A perhaps more direct way to get a hold of the result is to consider the  source
\beq
{\cal J}_{1+y}(\bar g,\tau)\equiv 
\exp\left\{-y\Dw_\tau\right\} {\cal J}\Big|_{{\cal J}={\cal J}_1}= 
( \bar g^{\mu\nu}e^{2(1-y)\tau} ,\,\lambda^I[\tau,y], \,A_\mu^A[\tau,y], \, m^a[\tau,y])\, ,
\label{J_1+y}
\eeq
which interpolates between ${\cal J}_1$ at $y=0$ and ${\cal J}_2$ at $y=1$. The advantage of using this interpolating source is readily seen when considering $\Gamma_{local}[\tau]$. One can indeed write
\beq
\Gamma_{local}[\tau]=\WW[{\cal J}_1]-\WW[{\cal J}_2]=-\int_0^1 dy \frac{d}{dy}\WW[{\cal J}_{1+y}]=\int d^4x\int_0^1 dy  \AA_\tau({\cal J}_{1+y})
\label{integral_y}
\eeq
where $\AA_\tau({\cal J}_{1+y})$ is just the Weyl anomaly of eq.~\eqref{eq_osborns_anomaly} computed for Lie parameter $\sigma=\tau$ on the background ${\cal J}_{1+y}$.
To compute both pieces in $\Gamma[\bar g,\tau]$ we must then first find ${\cal J}_{1+y}$. This is done by  solving a set of differential equations. Indeed, by its definition, ${\cal J}_{1+y}$  satisfies
\beq
\frac{d}{dy} {\cal J}_{1+y} = -\Delta_{\tau} {\cal J}_{1+y}
\eeq
which corresponds to a set of first order differential equations for its components. Given eq.~\eqref{deltasource} the solution is found by considering $\lambda^I$ first, $A^A_\mu$ second and $m^a$ third. We have
\beq
\frac{d}{dy} \lambda^I[\tau,y] = \tau B^I(\lambda[\tau,y])
\eeq
which, with initial condition $ \lambda^I[\tau,0]=\lambda^I(\mu)$, has  solution
\beq
\lambda^I[\tau,y] = \lambda^I(\mu e^{y\tau})
\eeq
This result is obvious given the definition of $\lambda^I[\tau,y]$ in eq.~\eqref{J_1+y}, but for the other sources the result will be less obvious. Consider now the vector field. One has
\beq
\frac{d}{dy} A_\mu^A[\tau,y]= \tau y B^IP_I^A \n_\mu\tau -\tau P_I^A(T_B\lambda)^I A^B_\mu[\tau,y]
\eeq
where $\lambda^I\equiv \lambda^I(\mu e^{y\tau})$  is  understood everywhere. Notice moreover that by the relation $B^IP_I^A=0$ only the homogeneous term survives. Thus, given the initial condition  $A_\mu^A[\tau,0]=0$,  the unique solution is $A_\mu[\tau,y]= 0$. This is an interesting and non-trivial result. It implies that 
  $\Gamma_{local}$ is not affected by anomaly terms involving the field strength of the external gauge fields, while 
$\Gamma_{non-local}$  is  independent of the correlation functions of the Noether currents $J_\mu^A$. We stress that this result depends on the choice $S^A=0$ and would not hold otherwise. As we saw in section  \ref{sec_ambiguity}, setting $S^A=0$ amounts to using the Ward identity eq.~\eqref{eq_naive_WI} to eliminate $\partial_\mu J_A^\mu$ in the expansion of $T$ in eq.~\eqref{eq_T_beta_S_t}. What our present argument shows, is that $J_A^\mu$ is eliminated altogether, including  the general case where operators are inserted at coinciding points and contact terms must be taken into consideration.

Consider finally $m^a$. Its Weyl transformation is somewhat intricate, and so is the differential equation for $m^a[\tau,y]$. The computation is considerably simplified by focussing instead on the ``covariant" quantity $\Pi^a[\tau,y]$. This is simply related to $m^a[\tau,y]$ (see eq.~\eqref{eq_Pi}) via the sources  we already computed, the metric $g^{\mu\nu}[\tau,y]\equiv \bar g^{\mu\nu}e^{2(1-y)\tau} $ and $\lambda^I[\tau,y]$. By eq.~\eqref{eq_pi_transformation} the equation it satisfies is 
\bea
\frac{\delta}{\delta y} \Pi^a [\tau,y]&=&-\tau\left\{ (2-\gamma)^a_b \Pi^b [\tau,y] +\gamma^a_{IJ} \Pi^{IJ} [\tau,y]\right \}\\
&=& -\tau\left\{(2-\gamma)^a_b \Pi^b [\tau,y] +  e^{2\tau y} \l 6C^a + \LL[\wt \eta^a] \r  \l \frac{1}{6}\br R  + \n^2\tau   -(\n\tau)^2\r\right \}\nl
\eea
where $\tilde \eta^a=\eta^a + \half  \theta^a_I{(U^{-1})}^I_J B^J$ and where, as before, $\lambda^I\equiv \lambda^I(\mu e^{y\tau})$ is understood everywhere. In the second line we have used the explicit expression for 
$\gamma^a_{IJ} \Pi^{IJ} [\tau,y]$, which is readily computed as this quantity purely depends on $\lambda^I$ and on the metric. Furthermore we have used its definition and the consistency conditions to rewrite the coefficient $\gamma^a_{IJ}$. This is a standard differential equation whose solution is formally written in terms of integrals involving the known functions on the right-hand side. The dependence on $\tau$ can then be made explicit by expanding the formal solution in a Taylor series in $\tau$. 

The structure of $\Pi^a$ is the main source of complication in the  computation of $\Gamma[\bar g,\tau]$ for general $\tau$. In the Appendix we give more details about  the general case. Here we shall instead focus on the specific dilaton field configurations respecting the ``on-shell condition"
\beq
R(\bar g^{\mu\nu}e^{2\tau})=e^{2\tau}\l  \br R  +6\left [\n^2\tau   -(\n\tau)^2\right ]\r =e^{2\tau}\l \br R  -6e^{\tau}\n^2e^{-\tau}\r =0\, 
\label{on-shell_g}
\eeq
which for the flat background $\bar g^{\mu\nu}=\eta^{\mu\nu}$ reduces to the massless Klein-Gordon equation for the ``canonical" dilaton $1+\phi\equiv e^{-\tau}$.
 The effective action for a dilaton satisfying the on-shell condition very roughly generates the correlators of $T$ for lightlike external momenta, though  the relation is more involved because of contact terms.  These configurations are interesting because they are precisely those that help  constraining the structure of the RG flow \cite{Luty:2012ww}.
Now, in the case of an on-shell dilaton, a remarkable simplification takes place: $\Pi^a [\tau,y]=\Pi^{IJ}[\tau,y]=0$. Indeed  one readily checks that for on-shell configurations  the boundary condition is $\Pi^a [\tau,0]=\Pi^{IJ}[\tau,0]=0$. Then, since the system of $\Pi[\tau,y]$'s satisfies a homogeneous differential equation (see eq.\eqref{eq_pi_transformation}), the solution vanishes identically. By the explicit form of $\Pi^a$ we thus have that on-shell and for a flat metric
\beq
\Pi^a[\tau,y]=0 \quad\longrightarrow\quad m^a[\tau,y]=e^{2(1-y)\tau}\left [y(1-y) \eta^a+y^2\frac{\theta_I^a}{2}B^I\right ] \Box \tau 
\eeq
where again all coefficients implicitly depend on $\tau$ and $y$ via $\lambda^I\equiv \lambda^I(\mu e^{y\tau})$. Notice that for $y=1$, relevant for the computation of $\Gamma_{non-local}$, the above result further simplifies to (all $\tau$ dependence now explicit)
\beq
 m^a[\tau,1]=\frac{\theta_I^a(\lambda(\mu e^{\tau}))}{2}B^I(\lambda(\mu e^{\tau}))\Box \tau\, .
 \label{m_tau}
\eeq
We have now all the ingredients to quickly evaluate the dilaton effective action in the on-shell case (The off-shell case is discussed in  appendix \ref{app_non_local}). We shall consider the local and non-local contributions separately.

\subsection{Computation of $\Gamma_{local}$}
\label{sec_local_interactions}

From eq.~\eqref{integral_y} we see that $\Gamma_{local}$ is linear in the anomaly. It thus consist of the addition of 5 terms, one for each of the contributions in eq.~\eqref{eq_Weyl_anomaly_Off_criticality}.
\beq
\Gamma_{local}=\Gamma^{\n^2 R}+\Gamma^{R^2}+
\Gamma^{W^2}+\Gamma^{E_4}+\Gamma^{F^2}\eeq

\begin{enumerate}

\item{$\Gamma^{\n^2 R}$}

This local contribution 
 can be obtained by dividing the generating functional into two pieces
\bea
\WW &=& \WW' -  {\cal F}_{\n^2 R}
\eea
where 
\bea
-\Delta_\sigma {\cal F}_{\n^2 R} &=& \int d^4 x \AA_\sigma^{\n^2 R}~.
\eea
while $\WW'$ is a modified action whose anomaly has the canonical form $\AA^{R^2}+
\AA^{W^2}+\AA^{E_4}+\AA^{F^2}$. The explicit expression for $ {\cal F}_{\n^2 R}$ is given in \eqref{eq_local_term_n2r}.
By the definition eq. \eqref{eq_diltaon_effective_action}, and by using eq.~\eqref{eq_local_term_n2r} we then simply have
\bea
\Gamma^{\n^2 R}[\gf, \tau]=-{\cal F}_{\n^2 R} [\JJ_1]
=-  \int d^4x\sqrt {-\bar g}~\wt d
\l
\frac {R[\br g]}{6}+\n^2\tau - \l \n\tau\r^2\r^2
\eea
where 
$\wt d$ is given by
\bea
\label{eq_tilde_d}
\wt d  &=&d + \half B^I U_I+
\frac 1 4   \wt S_{(IJ)}B ^I \wt B^J
-\eta^ak_a -\half \eta^a j_{aI} \wt B^I 
\eea
and we introduced the notation $\wt B^I = (U^{-1})^I_JB^J$. 
It is important that once we have extracted this piece from the generating functional, the remaining terms must be evaluated using $\WW'$, namely in a scheme where the generalized $\n^2 R$ anomaly vanishes.

The main result here is that $\Gamma^{\n^2 R}$ vanishes for dilaton configurations satisfying eq.~\eqref{on-shell_g}. As such this contribution  does not affect the discussion on the RG flow structure: that makes sense, since the local functional ${\cal F}_{\n^2 R}$ is arbitrary.

\item{$\Gamma^{R^2}$}

This  contribution is given by the integral of a quadratic form in the $\Pi's$. It is therefore proportional to the square of  $\br R  +6\left [\n^2\tau   -(\n\tau)^2\right ]$, and therefore trivially vanishes for on-shell dilaton configurations.

\item{$\Gamma^{W^2}$}

The contribution from $\AA^{W^2}$ is easily integrated: $\sqrt g W^2$ is Weyl invariant, so  that  the only dependence on $\tau$ and $y$  comes from the coefficient function $c(\lambda)$. We find
\bea
\Gamma^{W^2}[\gf, \tau]
=
- \int dx \sqrt {- \gf} ~ C(\lambda(\mu),\tau) W^2[\gf]
\eea
where $C(\lambda(\mu),\tau) = \int_{\mu}^{\mu e^\tau} c(\lambda(\bar \mu))d\ln \bar \mu$.
This contribution vanishes in a flat metric background.

\item{$\Gamma^{E_4}+\Gamma^{F^2}$}

We group these two  contributions, since $\AA^{E_4}$ and $\AA^{F^2}$ are related by the Wess-Zumino consistency condition. Notice however that since $A_\mu^A[\tau,y]=0$, the gauge  field strength vanishes and $\AA^{F^2}$ reduces to the terms proportional to $P_I^A$. 
We find
\bea
\Gamma^{E_4}[\gf,\tau]
=
 \int d^4x \sqrt{-\gf}
&&\Bigg(
 A(\lambda(\mu),\tau)
  E_4[\gf]\nl
  &&+  \wt a(\lambda(e^\tau\mu)) \l  4 G^{\mu\nu}[\gf]\n_\mu\tau\n_\nu\tau
 - 4\n^2\tau \n_\mu\tau\n^\mu\tau +2\l \n_\mu\tau\n^\mu\tau\r ^2
\r   \nl 
&&-  \LL[\wt a] (\lambda(e^\tau\mu))   \l \n_\mu\tau\n^\mu\tau\r ^2+\ldots
\label{E_4action}
\eea
where 
$A(\lambda(\mu),\tau) = \int_{\mu}^{\mu e^\tau} a(\lambda(\bar \mu))d\ln \bar \mu$, while 
the dots stand for additional terms  of order $O(B)^2$ and  proportional to $\br R  +6\left [\n^2\tau   -(\n\tau)^2\right ]$. These additional terms therefore vanish on-shell.

Notice that  eq.~\eqref{mudmua}  implies $8 \LL[\tilde a]= \chi_{IJ}^gB^I B^J=O(B^2)$. Therefore, close  to the fixed point, where we can use $B^I$ as a small expansion parameter, and focussing on a flat metric, the above formula reduces to 
\bea
\label{eq_beta_b_interactions_near_fp}
\Gamma^{E_4}[\eta,\tau]
&=&
 \tilde a
 \int d^4x \l- 4\n^2\tau \n_\mu\tau\n^\mu\tau +2\l \n_\mu\tau\n^\mu\tau\r ^2\r
 + O(B^2) 
\eea
This has precisely the form of the Wess-Zumino term at the fixed point\cite{Komargodski:2011vj}: the non-trivial result is that the corrections begin only at order $(B^I)^2$. 

 \end{enumerate}
Let us summarize:  for flat background metric $\bar g^{\mu\nu}=\eta^{\mu\nu}$ and for $\tau$  satisfying the on-shell condition $\Box e^{-\tau}=0$, the local contribution to the  effective action is controlled by the anomaly coefficient $\tilde a$ and reduces  to the second and third lines of eq.~\eqref{E_4action}.

\subsection{Computation of $\Gamma_{non-local}$}
\label{sec_non_local}

As long as we are not interested in correlators involving $T_{\mu\nu}$ we can set $\bar g^{\mu\nu}=\eta^{\mu\nu}$. Using the results in section \ref{dilaton_action}, we have
\beq
\Gamma_{non-local}=\WW[{\cal J}_2]= \WW[\lambda[\tau,1],m[\tau,1]]
\eeq
where, with a slight abuse of notation, we have dropped the metric and gauge field as one is flat and the other vanishes. By writing 
\beq
\label{eq_non_local_general}
\WW[\lambda[\tau,1],m[\tau,1]]=\exp\left \{\int d^4 x \left [(\lambda[\tau,1]-\lambda^I(\mu))\frac{\delta}{\delta \bar \lambda^I(x)}+m^a[\tau,1]\frac{\delta}{\delta \bar m^a(x)}\right ]\right\} \WW[\bar \lambda,\bar m]\Big\vert_{\bar \lambda=\lambda(\mu), \,\, \bar m=0}
\eeq
and by using the functional correspondence between derivatives and operators, the have that the $\tau$ dependence of $
\Gamma_{non-local}$ is effectively generated by adding to the lagrangian of the QFT an effective interaction (we use $[\lambda[\tau,1]=\lambda(\mu e^\tau)$)
\beq
{\cal L}_{eff}= (\lambda^I(\mu e^\tau)-\lambda^I(\mu)) {\cal O}_I+(m^a[\tau,1]){\cal O}_a\, .
\eeq
In the case of an on-shell dilaton the explicit result is
\beq
\label{lefftau}
{\cal L}_{eff}=(\lambda^I(\mu e^\tau)-\lambda^I(\mu)) {\cal O}_I+\frac{\theta_I^a(\lambda(\mu e^{\tau}))}{2}B^I(\lambda(\mu e^{\tau}))\Box \tau {\cal O}_a\, .
\eeq
where of course the composite operators are also renormalized at the scale $\mu$. Because of the piece  proportional to $\theta_I^a$, this result corrects  the naive expectation according to which in a QFT with purely marginal deformations the effective coupling to  a background dilaton is simply obtained by promoting $\lambda(\mu)$ to $\lambda(\mu e^{\tau})$. That would for instance be automatically  true in the absence of dimension 2 scalars. However, we have seen before that even in the presence of dimension 2 operators a scheme to define composite operators exists where $\theta_I^a=0$. In such a scheme the form of the effective dilaton  interaction would respect the naive expectation. Notice that the operator redefinition generated by the source reparametrization in eq.~\eqref{reparsources}, reduces to a simple operator shift, as described in eq.~\eqref{oshift}, only when operators are inserted at separated points. When considering insertions at coinciding points the operator mapping is made more involved by the presence of contact terms. 
In view of that, one should not be worried if the second term in eq.~\eqref{lefftau} cannot be naively absorbed by the first through a simple operator shift.

We should however stress that the simple result in eq.~\eqref{lefftau} relies on two other ingredients. First, it relies on the choice $S^A=0$ to fix the freedom in defining the RG flow. This choice is equivalent to using the Ward identity to rewrite $\partial_\mu J^\mu_A$  in terms of ${\cal O}_I$ and ${\cal O}_a$. Secondly, and more importantly,  eq.~\eqref{lefftau} is only valid for on-shell dilatons. Without that assumption there would be new genuine contributions basically related to the  existence of the additional non-minimal operators $\sqrt g R(g) {\cal O}_a$ coupling the QFT to gravity.

For the purpose of the discussion in the next section, it is useful to write  the lowest order contributions to $\Gamma_{non-local}$ in an expansion in the canonical dilaton $\phi$
\bea
e^{-\tau}=1+\phi~.
\label{tauphi}
\eea  
for which  the on-shell condition is  $\n^2\phi=0$. Using the expansions  
$\tau=-\phi+\half \phi^2-\ldots$ and $\n^2\tau=-(1-\phi)\n^2\phi +(\n\phi)^2+\ldots$ we find
\bea
\label{eq_non_local_operator_phi}
\Gamma_{non-local}[\eta,\phi]&=&:\exp \Bigg\{\int d^4x \Bigg( -\phi ~B^I \frac{\delta}{\delta \lambda^I(x)}
\nl&&~~~~~~~~~~~~~~~~~~~
+\frac{ \phi^2}{2} \l  B^J\l \delta_J^I + \d_J B^I \r  \frac{\delta}{\delta \lambda^I(x)}
+\half B^J\theta_J^a\n^2  \frac{\delta}{\delta m^a(x)}\r 
\nl&&~~~~~~~~~~~~~~~~~~~ 
+O(\phi^3)
\Bigg)\Bigg\}:\WW \Big|
\eea
where by the $: \, :$ we mean that the functional derivatives do not act on their coefficients.
As a check of the consistency of our result notice that  the term proportional to $\phi^2$ is given by 
\bea
\int d^4x \frac{\phi^2}{2} B^J\l \delta_J^I + \d_J B^I \r \left [\wt \OO_I\right]
\eea
where $\wt \OO_I$ is the scheme independent dimension 4 operator defined in eq. \eqref{eq_scheme_independent_OI}.  Also consistently with that: thanks to $\Box \phi=0$,  $\OO_I$ and $\wt \OO_I$ make no difference in the term linear in $\phi$.

\subsection{Correlators of $T$ and the constraints on the RG flow}
\label{sec_LPR}
The  constraint on the RG flow asymptotics discussed in section \eqref{sec_gradient_flow} can be alternatively  derived
by studying  the specific  combination of correlators of $T$ that corresponds to the $2\to 2$ scattering amplitude of a background on-shell dilaton.
 This  approach is at the basis of the 
proof of the $a$-theorem in ref \cite{Komargodski:2011vj} and  was already followed in ref. \cite{Luty:2012ww} to constrain the 
RG flow asymptotics. This section has a  twofold aim. On one hand we would like to use the results of the previous section to fill in some details that where not fully developed in ref. \cite{Luty:2012ww}.
These concern the role of multiple insertions of $T$, and  the issues of scheme dependence and operator mixing. In the end these issues affect only subleading contributions and so they do not   alter the proof  in ref. \cite{Luty:2012ww} as, under the assumption of perturbativity, that only relies on the leading order scattering amplitude. However, with a complete control of the scattering amplitude, the relation with the consistency condition approach of refs.\cite{Osborn:1989td,Jack:1990eb,Osborn:1991gm} will be more clear. That is our second aim. 

The  idea is to study specific combinations of correlators of $T$ that can be directy interpreted as the $2\to 2$ scattering amplitude of the background dilaton field $\phi$ defined in eq.~(\ref{tauphi})
\beq
(2\pi)^4 \delta(p_1+\dots+p_4)A(p_1,p_2,p_3,p_4)=
\frac{\delta}{\delta \phi(p_1)}
\frac{\delta}{\delta \phi(p_2)}
\frac{\delta}{\delta \phi(p_3)}
\frac{\delta}{\delta \phi(p_4)} \WW [\JJ_1] \Big|_{\gf=\eta,\phi=0}
\eeq
Notice that since 
\beq
\frac{\delta }{\delta \phi}=-e^{\tau}\frac{\delta}{\delta \tau}
\eeq
the  amplitude is a combination of 4-, 3- and 2-point functions 
\bea
A(p_1, \ldots, p_4) &=&-i \TO{ T(p_1) T(p_2) T(p_3) T(p_4)} 
\nonumber\\
&& 
- \l \TO{ T(p_1 + p_2) T(p_3) T(p_4)} + \mbox{permutations}\r 
\nonumber\\
&&
+i \l \TO{ T(p_1 + p_2) T(p_3 + p_4)} +  \mbox{permutations}\r 
\nonumber\\
&& +i\l  \TO{ T(p_1 + p_2 + p_3) T(p_4)}  + \mbox{permutations}\r  \,.
\label{amplitude}
\eea Notice that, for generic kinematics, the correlators of $T$ require renormalization. As a result of that, these correlators are generically $\mu$ dependent. An equivalent statement is that the dilaton effective action for a generic $\phi$ is $\mu$ dependent. As discussed in section 2.2.1 this dependence is fully controlled by the integral of the anomaly for a constant variation parameter $\sigma= {\mathrm {const}}$. Now, it turns out that, for a pure dilaton background $g_{\mu\nu}=\eta_{\mu\nu}(1+\phi)^2$
 satisfying the ``on-shell" condition
\beq
R(e^{-2\tau}\eta_{\mu\nu})=e^{3\tau}\Box e^{-\tau}=(1+\phi)^{-3}\Box \phi=0
\label{on-shell}
\eeq
the anomaly of eq.~(\ref{eq_osborns_anomaly}) integrates to zero. Indeed, in a pure dilaton background ($\lambda^I={\mathrm{const}}$, $A_\mu^A=m^a=0$) the only  terms to consider are those involving just the metric:  $E_4$ integrates to zero over asymptotically flat space, 
$\sqrt{- g} W^2(g)$ vanishes for conformally flat metrics, while the on-shell condition \eqref{on-shell} eliminates the $R^2$ term. The scattering amplitudes for on-shell dilatons are thus automatically finite, that is they are RG independent.

 The same conclusion can be obtained from the power counting  analysis in ref. \cite{Luty:2012ww}, from which  one deduces that for an on-shell dilaton background all counterterms vanish except for a cosmological constant term $\frac{\Lambda}{4!}(1+\phi)^4$.
For $m^a\not =0$  the cosmological term would logarithmically depend on $\mu$. This dependence is associated with the $\Pi^a\Pi^b$ terms in the anomaly.  However, for the case $m^a=0$ we are interested in  there is just a quartic divergence: $\Lambda$ is a $\mu$ independent constant, that we may in principle even set to zero. Indeed eq \eqref{eq_osborns_anomaly} corresponds to the choice $\Lambda=0$.

As a consequence of the above discussion, on dimensional grounds, the scattering amplitude, takes the form
\beq
A(s,t)=s^2 F(s/\mu^2,t/\mu^2,\lambda(\mu)) + \Lambda
\eeq
with $F$ an RG invariant function
\beq
\left (\mu\frac{\partial }{\partial\mu} + B^I\frac{\partial}{\partial \lambda^I}\right ) F(s/\mu^2,t/\mu^2,\lambda(\mu))=0\, .
\label{rgF}
\eeq
Notice  that, since the dilaton is a flavor singlet source, $F$ must be invariant under the background $G_F$:  in   eq.~(\ref{rgF}) we can equally well  use $B^I$ or   $\beta^I$.

The constraint on the flow is obtained by considering a dispersion relation for the forward scattering amplitude $A(s,t=0)$ \cite{Komargodski:2011vj,Luty:2012ww}.
In principle, given the  kinematics $(p_i^2=0, t=0)$, one may be concerned about the IR finiteness of the amplitude.
While we believe it should be possible to carefully study the conditions for IR finiteness by performing an operator expansion analysis, in the present study we shall content ourselves by assuming the amplitude is finite.
There are different reasons to believe that must be the case. One is that, as it will become clear below,  $A(s,t=0)$ appears to provide a concrete ``on-shell" scheme to define the quantity $\tilde a$ that emerged from the study of the consistency conditions. It seems hard to believe that happens just by chance. Another, maybe weaker, indication is associated with the explicit form of $A(s,t=0)$, when expanded in powers of the $\beta$-function. As we shall  discuss below, at  the leading $\beta^2$ order, the amplitude is determined by the two point functions of operators $\wt O_I$
with dimension near $4$, and is manifestly IR finite. The next-to-leading order $\sim \beta^3$ is determined by 3-point functions of such operators, which at lowest order in $\beta$ can be computed in the original unperturbed CFT. Here again, the explicit computations of CFT 3-point in momentum space \cite{Bzowski:2013sza}, allows to rule out IR singularities. According to this reasoning IR singularities could only arise beyond the order $\beta^4$. While this seems difficult to believe, a dedicated analysis seems to be needed to rule out this possibility. We leave such analysis for future work.

\begin{figure}[htbp]
\begin{center}
		\includegraphics[height=80pt]{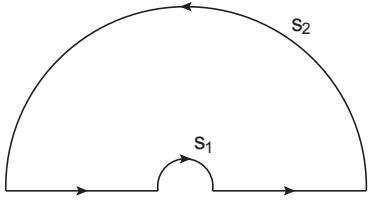}
		
	\caption{The contour $C$ in the complex $s$ plane. }
\end{center}
	\label{fig:contour}
\end{figure}

Let us now go back to the forward amplitude. It is useful to parametrize it as  
\beq
A(s,0)=s^2F(s/\mu^2,0,\lambda(\mu)) + \Lambda\equiv -8 s^2\alpha(s)+\Lambda
\label{Aalpha}
\eeq
such that  the positivity constraint imposed by unitarity becomes
\beq
{\mathrm{Im}}A(s,0)\geq 0    \qquad\Longrightarrow\qquad {\mathrm{Im}}\alpha \leq 0
\label{ImA}
\eeq
Notice that, by the results of sections \ref{sec_local_interactions}-\ref{sec_non_local}, eq.~\eqref{E_4action} in particular,  at a conformally invariant fixed point,  $ \alpha$ coincides with the anomaly coefficient $a$. Away from criticality, using the $\mu$ independence of $A$, we can also write
\beq
-8 \alpha(s)= F(1,0,\lambda(\sqrt s)), 
\eeq
 a  finite function of the running couplings. The dispersion relation corresponds to the Cauchy integral relation
 \beq
 \oint_C \frac{A(s,0)}{s^3} ds =0
 \label{cauchy}
 \eeq
for the contour $C$ shown in figure \ref{fig:contour}. By using crossing  $A(s,0)=A(-s,0)$ and ``hermiticity" $A(s,0)^*=A(s^*,0)$, and by defining the ``average" amplitude 
\beq
\bar \alpha(s) =\frac{1}{\pi} \int_0^\pi \alpha(se^{i\theta}) d\theta
\label{alphabar}
\eeq
eq.~(\ref{cauchy}) becomes \cite{Luty:2012ww}
\beq
\bar\alpha(s_2)-\bar\alpha(s_1)=\frac{2}{\pi}\int_{s_1}^{s_2} \frac{ds}{s} (-{\mathrm{Im}} \alpha(s)) \geq 0
\label{dispersion}
\eeq
Notice that by crossing and hermiticity, $\bar \alpha$ is a real quantity.
Notice also that the cosmological term, being analytic over the whole complex plane automatically gives no contribution to the dispersion relation.

We can now use the results from our study of the local Callan-Symanzik equation to elucidate both sides of eq.~\eqref{dispersion}. 
Consider the left-hand side first. The split of the dilaton effective action into a local and non-local contribution corresponds to a similar splitting for the dilaton amplitude $\alpha=\alpha_{loc}+\alpha_{non-loc}$. The results of  the two previous sections  imply
\beq
\alpha_{loc}= \tilde a(\mu) +O(B^2) \qquad\qquad \alpha_{non-loc}=O(B^2)
\eeq
from which, using the $\mu$ independence of $\alpha$, we deduce $\bar \alpha$ satisfies
\beq
\bar\alpha(s) =\tilde a(\sqrt s)+O(B^2)
\label{alpha-a}
\eeq
This  relation is sufficient to conclude that there exists a choice of scheme where $\bar\alpha(s) =\tilde a(\sqrt s)$. Indeed adding to $\WW$  the local term
\beq
\frac{c_{IJ}}{2} \sqrt g G^{\mu\nu}\n_\mu\lambda^I\n_\nu \lambda^J
\eeq
does not affect the dilaton amplitude, as that is computed at $\n_\mu\lambda^I=0$, but modifies $\tilde a$ and $\chi_{IJ}^g$ according to
\beq
\tilde a \to \tilde a +B^IB^J c_{IJ}\qquad\qquad \chi_{IJ}^g \to  \chi_{IJ}^g +{\cal L}(c_{IJ})\, .
\eeq
The first equation, together with eq.~(\ref{alpha-a}), implies a $c_{IJ}$ with regular dependence on $\lambda^I$ can be chosen such that  $\bar\alpha(s) =\tilde a(\sqrt s)$.
\begin{figure}[htbp]
\begin{center}
		\includegraphics[height=80pt]{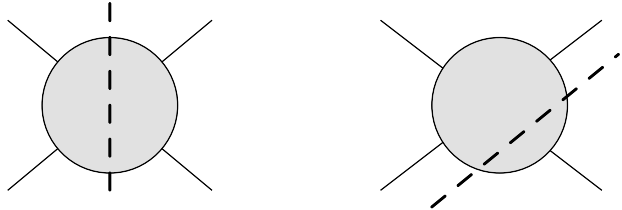}
	\caption{The 2-2 and 3-1 cuts of the on-shell dilaton scattering amplitude.}
	\label{fig:cuts}
\end{center}
\end{figure}
Consider now the right-hand side of eq.~(\ref{dispersion}). The imaginary part of the amplitude is obviously only affected by the non-local part of the dilaton action. We must thus expand $\Gamma_{non-loc}$ to fourth order in $\phi$. Notice first of all, as it may also seem obvious, that only 2-2 cuts   contribute\footnote{Indeed this is necessary to establish eq.(\ref{ImA}), as 2-2 cuts are manifestly positive while 3-1 cuts are not manifestly positive.} if the amplitude is assumed to be finite for external momenta on the lightcone:
 3-1 cuts would expectedly be associated with singularities 
at $p_i^2=0$.  The absence of 3-1 cuts physically  corresponds to the fact that a background massless dilaton cannot decay to QFT states. This last statement can also be checked by noticing that the contribution from   $\Gamma_{non-local}$ to the dilaton 2-point function vanishes on-shell. 

\begin{figure}[htbp]
\begin{center}
		\includegraphics[height=80pt]{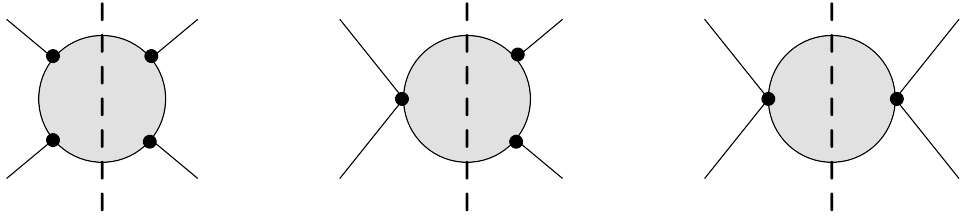}
	\caption{The different configurations for the diagrams with 2-2 cut.}
	\label{fig:amplitude}
\end{center}
\end{figure}

Now, since only 2-2 cuts contribute to the imaginary part, we must consider terms where at most two $\phi$'s are at coinciding point, as shown in the fig.~\eqref{fig:amplitude}.
 The contributions with at most two coinciding $\phi$'s are determined by the $O(\phi^4)$ term in the expansion of eq.~(\ref{eq_non_local_operator_phi}). These contributions can be written in terms of ``Feynman rules" where the building blocks are 2- 3- and 4- point correlators of ${\cal O}_I$ and ${\cal O}_a$.   Inserting a complete set of states $|\Psi\rangle$ in the cut, the imaginary part is conveniently written as
 \beq
-{\mathrm{Im}}\,{\alpha}(s)\,=\,\frac{1}{16 s^2}\sum_\Psi\,(2\pi)^4\delta^4(p_\Psi -p_1-p_2)B^IB^J{\cal M}_J(\Psi)^*{\cal M}_I(\Psi)
\eeq
with the matrix elements defined as
\beq
B^I{\cal M}_I(\Psi)\equiv B^I\langle\Psi | \left [(\delta_I^K+\partial_IB^K)\wt {\cal O}_K(0)+B^K{\cal O}_{IK}(p_1-p_2)\right ]|0\rangle
\label{matrixelement}
\eeq
where we used the ``scheme independent" dimension 4 operator $\wt \OO_I$ defined in eq. \eqref{eq_scheme_independent_OI}, and defined
\beq
{\cal O}_{IK}(p_1-p_2)\equiv \int d^4y e^{-i(p_1-p_2)y/2}{\bf T}\left ({\cal O}_I(y){\cal O}_K(-y)\right )\, .
\eeq
$p_1$ and $p_2$ are the momenta of the two incoming dilatons, so that $(p_1+p_2)^2=s$.
The matrix element $B^I \mathcal M_I(\Psi)$ describes the probability amplitude for two incoming dilatons to be converted into the state $|\Psi\rangle$. The first two terms in eq.~(\ref{matrixelement}) correspond to two dilatons absorbed at coinciding points (pure $\ell=0$-wave) while the third corresponds to insertions at non-coinciding points, and thus involves all higher partial waves $\ell\geq 0$. 

One can thus define a positive metric $G_{IJ}$ such that 
\bea
-{\mathrm{Im}}\,{\alpha}(s)\,&=&\,B^IB^J G_{IJ}\\
G_{IJ}&=&\frac{1}{16 s^2}\sum_\Psi\,(2\pi)^4\delta^4(p_\Psi -p_1-p_2){\cal M}_J(\Psi)^*{\cal M}_I(\Psi)
\eea
In the above  equation, by the $\mu$ independence of the amplitude, the couplings and the composite operators can be conveniently renormalized at $\mu=\sqrt s$.
Plugging this result into  eq.~(\ref{dispersion}) and comparing to eq.~(\ref{integralbound}) one concludes that, in the scheme  $\tilde a(\sqrt s)=\bar\alpha(s)$,
\beq
\chi^g_{IJ}=\frac{32}{\pi} G_{IJ} +\Delta_{IJ}
\label{gGDelta}
\eeq
where $\Delta_{IJ}$ satisfies $B^IB^J\Delta_{IJ}=0$, while $G_{IJ}$ is manifestly positive definite. The positive matrix $G_{IJ}$ can be viewed as the 4D analogue of Zamolodchikov's metric for 2D RG flows
\beq
G_{IJ}^{2D}\equiv \frac{1}{p^2}\sum_\Psi (2\pi)^2\delta^2(p_\Psi -p) \langle 0 | {\cal O}_I(0)|\Psi\rangle \langle\Psi | {\cal O}_J(0)|0\rangle\, .
\eeq
With the benefit of hindsight we can now better appreciate the difference between the 2D and 4D cases. In the first case the RG flow is controlled by the 2-point correlator of $T$, while in the second a specific combination of 2-, 3-, and 4-point correlators is the relevant object. Without the dilaton scattering amplitude as a guideline it would not have been obvious how to assemble these correlators in order to construct  $G_{IJ}$. Of course the approach we followed in this paper is bound to the study of near marginal deformations
where both $B^I$ and $\partial_I B^J$ are treated as small perturbations. In that case $G_{IJ}$ is dominated by the first term in eq.~(\ref{matrixelement}) and takes the same 
2-point function structure fo the 2D case. That is the result discussed in ref. \cite{Luty:2012ww}. Ideally one could however conceive of extending  eq.~(\ref{matrixelement}) beyond perturbation theory including all scalar operators in the theory \cite{osborn}. Unitarity would then dictate the evolution of $\bar \alpha$ with energy is controlled by an infinite dimensional positive metric constricted in analogy with $G_{IJ}$.

We want to conclude with a comment concerning parity violation and $\epsilon^{\mu\nu\rho\sigma}$ terms in the anomaly. In this paper we have disregarded them in order to simplify the discussion on the structure of the anomaly. However it is rather clear that their presence does not affect the derivation of the effective action for the dilaton, and the discussion about RG flow based on it. This is readily seen by considering in turn $\Gamma_{local}$ and $\Gamma_{non-local}$. The former is a local action involving 4 derivatives and any power of a scalar field $\tau$: by Bose symmetry it is evident that one cannot write down any term involving $\epsilon^{\mu\nu\rho\sigma}$. The latter is totally determined by the Weyl tranformation properties of the sources, which as we noticed  in section \ref{sec_local_CS}, is not affected by parity violation. Therefore the discussion of RG flow asymptotics is not affected by parity violation and, consequently, by mixed flavor-gravity anomalies.

\section{Conclusions}
\label{sec_conclusions}

Osborn's original paper \cite{Osborn:1991gm} on the local RG outlined a beautiful formalism to shed light on the structure of RG flows,  independent of details of the underlying theory. The present  can be largely considered as a corollary to that classic paper, where we obtained the following results: 
\begin{itemize}
\item We introduced 
the ``covariant"  objects $\Pi^a$ and $\Pi^{IJ}$ whose Weyl variations do not involve derivatives of the Lie parameter. These objects are essential in all applications of the local RG, from the  construction of manifestly consistent Weyl anomalies to the computation of the effective action for a background dilaton.
\item We showed that most of the consistency conditions for the Weyl anomaly  can be explicitly solved and  that the anomaly can be reformulated in a manifestly consistent form, with only 3 non-trivial consistency conditions remaining. A crucial step in that procedure was the isolation of the scheme dependent terms in the anomaly, that is terms that correspond to the variation of a local functional. That  allowed to write most consistency conditions as algebraic equations as opposed to differential equations. 
We believe this new formulation of the Weyl anomaly represents a significant simplification over the original discussion in ref. \cite{Osborn:1991gm}, providing focus on the genuinely non-trivial consistency conditions.
\item Using the full set of consistency conditions, in particular those involving the background flavor gauge field strengths, we derived a general gradient flow formula for the $\beta$-function, eq.~\eqref{our gradient flow}. This equation implies a certain combination of anomaly coefficients  $\tilde a=a+w_IB^I/8$ is stationary at fixed points. It turns out this is precisely the quantity that decreases monotonically when flowing towards the IR. Therefore maxima and minima of $\tilde a$ respectively correspond to UV and IR attractive fixed points. Another corollary of this result is that the $E_4$ anomaly coefficient $a$ is stationary on a manifold of fixed point.
\item We established the monotonicity of the RG flow of $\tilde a$, under the condition that the RG trajectory is bound to a neighbourhood of a CFT, where the $\beta$-function and the anomalous dimensions can be treated as small perturbations. These quantities are indeed the expansion parameters in all our computations. Our result evidently does not rely on the  original CFT being free.

\end{itemize}
We then  related the approach to gain insight on RG flows based on Weyl consistency conditions to the approach based on  the background dilaton trick of Komargodski and Schwimmer \cite{Komargodski:2011vj,Luty:2012ww}. Our study consists of the following steps and results:

\begin{itemize}
\item We derived a formal expression for the generating functional of the correlators of the energy momentum trace $T$: the  effective action for  a background dilaton $\tau$. This action consists of  two contributions. The first is local and determined by  the Weyl anomaly. For on-shell dilaton configurations  the result is fully determined by the $E_4$ anomaly term and shown in eq.~\eqref{E_4action}. A consequence of our result is that, up to ${\cal O}(B^2)$ in the $\beta$-function $B^I$, the forward dilaton scattering amplitude at energy $\sqrt s$ is controlled by  $\tilde a(\sqrt s)$, the same crucial quantity describing the gradient flow equation. This result was essentially derived already in ref. \cite{Fortin:2012hn}, though, we think, without analyzing the relevance of the on-shell condition.

The second contribution to the dilaton effective action is non-local and associated with the expansion of $T$ in terms of  a complete basis of operators, also including the effects of  multiple insertions at the same point.  Here the main result is that, for an on-shell dilaton, there exists  a suitable ``scheme" such that the action is simply generated by making the formal substitution $\lambda(\mu)^I\to \lambda^I(\mu e^\tau)$. On one hand the choice of scheme concerns the mixing between dimension 4-scalars ${\cal O}^I$ and operators of the form  $\Box {\cal O}^a$, with ${\cal O}^a$  dimension 2 scalars. On the other, it concerns the systematic use of flavor Ward identities to substitute the divergence of currents $\partial_\mu J^\mu_A$  in the correlators. That procedure corresponds to the freedom to define the Weyl operator such that $S^A=0$, and such that the $\beta$-function is the ``physical" one, $B^I$. We stress that, aside these technical scheme issues, the on-shellness of the background dilaton is the key to the simple result. In practice the on-shell condition beautifully filters out interactions (and related complications) associated with improvement terms. This property was already the key to the analyses in refs. \cite{Komargodski:2011vj,Komargodski:2011xv,Luty:2012ww}.

\item We used the effective action to study the   forward dilaton scattering amplitude. We showed that there exists a scheme where the  reduced forward amplitude $\bar \alpha(s)$, defined in eqs.~\eqref{Aalpha}\eqref{alphabar},  equals the quantity $\tilde a(\sqrt s)$ appearing in the study of Wess-Zumino consistency conditions \cite{Jack:1990eb,Osborn:1991gm}.
That scheme freedom is associated with the possibility to add   to the action a  local and finite functional of the sources. We then applied the optical theorem to show that, within this scheme and for a unitary theory, the matrix $\chi^g_{IJ}$ controlling the flow of $\tilde a$, essentially\footnote{``Essentially" is here used in the sense specified by eq.~\eqref{gGDelta}: a possible difference $\Delta_{IJ}$ would necessarily be ``orthogonal" to the $\beta$-funtion vector $B^I$ and not play any role.}   coincides with a positive definite metric in coupling space $G_{IJ}$. The latter metric is explicitly written in terms of matrix elements involving 2, 3- and 4-point correlators of the operators ${\cal O}_I$ that drive the RG flow. In practice the use of the dilaton scattering amplitude allows to identify the 4D analogue of the Zamolodchikov metric of 2D-QFT.

\end{itemize}

There are other directions in which one might proceed in the study of the local Callan-Symanzik equation and of the  dilaton effective action.
One obvious and straightforward exercise is to study the case of parity breaking QFTs allowing for mixed gravity-flavor anomalies. As we already mentioned, parity violation will not affect the discussion on the dilaton scattering amplitude and the RG flow of $\tilde a$. However it is not excluded that additional non trivial consistency conditions will  appear
involving the parity breaking coefficients. A less straightforward generalization from the methodological point of view is the one to supersymmetric field theories. Again, it is not excluded that additional and important consistency constraints will appear.  Supersymmetric RG flow lives on a (stable) submanifold of general QFT, where additional, and perhaps stronger, irreversibility constraint may apply. Ref.~\cite{Buican:2011ty} conjectures the existence of one such constraint valid in some specific cases. It is rather clear the local Callan-Symanzik equation offers a systematic methodology to hunt for all the possible constraints.

 \section*{Acknowledgments}

We would like to thank J.F. Fortin, M.A. Luty, H. Osborn, S. Rychkov and especially A. Monin for very useful discussions.
This work was supported by the Swiss National Science Foundation under grants 200020-138131 and 200020-150060.

\section*{Appendix}
\addtocontents{toc}{\protect\setcounter{tocdepth}{1}}

\appendix

\section{Definitions and useful equations}
\label{app_notations}
\subsection{Notations}
The generator of the local CS symmetry:
\bea
\Dw_\sigma&=&\Delta^g_\sigma-\Delta^\beta_\sigma
\eea
where
\bea
\Delta^g_\sigma = \int d^4x&\Bigg[&2\sigma ~g^{\mu\nu} \frac{\delta}{\delta g^{\mu\nu}(x)} \Bigg]\nl
\Delta^\beta_\sigma(x) = \int d^4x&\Bigg[&\sigma\l  \beta^I \frac{\delta}{\delta \lambda^I(x)}+    \rho_I^A  \n_\mu \lambda^I \frac{\delta}{\delta A^A_\mu(x)}\r  - \n_\mu\sigma  \l S^A  \frac{\delta}{\delta A^A_\mu(x)}\r  \nl
&& -\sigma\l m^b\l 2\delta_b^a  - \br \gamma_b^a\r 
+ C^a R+  D_I^a\n^2 \lambda^I + \half E_{IJ}^a\n_\mu \lambda^I \n^\mu \lambda ^J\r\frac{\delta}{\delta m^a(x)} \nl
&&+ \n_\mu\sigma \l \theta_I^a \n^\mu \lambda^I \frac{\delta}{\delta m^a(x)} \r
- \n^2 \sigma \l \eta^a \frac{\delta}{\delta m^a(x)} \r\Bigg]
\eea
Non-ambiguous functions in the local CS equation
\bea
B^I= \beta^I-\l S^AT^A\lambda\r^I\qquad
\gamma_b^a= \br\gamma_b^a- \l S^AT^A\r_b^a\qquad P_I^A= \rho_I^A +\d_IS^A
\eea
Notations appearing in the dilaton effective action
\bea
\label{eq_tilde_B_eta}
\wt B^I =  (U^{-1})^I_J B^J\qquad \wt \eta^a=\eta ^a + \half \theta^a_I\wt B^I
\eea
Useful anomalous dimension matrices
\bea
\label{eq_anomalous_dimensions_app}
\gamma^I_J&=& \d_J B^I + P_I^A(T_A \lambda^J)\nl
\gamma^K_{IJ}&=&\l U^{-1}\r^K_L\l \d_{(I} \gamma_{J)}^L + P_{(I}^A (T_A)_{J)}^L\r \nl
\gamma_{IJ}^{~a}&=&\half \l E_{IJ}^a + \theta^a_K\gamma^K_{IJ}\r \nl
\gamma_{IJ}^{KL}&=&B^{(K}\gamma^{L)}_{IJ} 
\nl
\gamma^B_A&=&P_J^B (T_A\lambda)^J\nl
U^I_J&=& \delta^I_J + \d_J B^I +\half P_I^A(T_A \lambda^J) 
\eea
Useful functions of the sources
\bea
 \Lambda^I&=&\l U^{-1}\r ^I_J\l  \n^2\lambda^J +\frac 1 6  B^J R \r \nl
\Pi^{IJ}&=&\n_\mu \lambda^I\n^\mu \lambda^J -B^{(I}\Lambda^{J)}
\nl
\Pi^a&=& m^a -\eta ^a \frac R 6 - \half \theta^a_I \Lambda^I
\nl
\Gamma^{\mu\nu}&=&G^{\mu\nu} + \frac R 6  g^{\mu\nu}\nl
\Omega^{IJK}&=&\l \Pi^{IJ} +  \half B^{(I} \Lambda^{J)}\r\Lambda ^K \nl
\Xi^{IJ}_\sigma&=&\Lambda^I \l 2\n_\mu\sigma \n^\mu\lambda^J - \sigma\gamma^J_{KL}\Pi^{KL} \r~. 
\eea

\subsection{Lie derivatives}
\label{app_Lie_derivative}
We use $\LL$ to denote a Lie derivative along a direction in parameter space defined by the RG flow.
This derivative satisfies the following definitions and relations: 
\bea
\LL[Y]&=&B^I\d_I Y\nl
\LL [Y_I^J]&=&B^K\d_K Y_I^J + \gamma_I^K Y_K^J- \gamma_K^J Y^K_I\nl
\LL [Y_{AI}]&=& B^J\d_J Y_{AI} +\gamma_I^J Y_{AJ}+ \gamma^B_A Y_{BI}\nl
B^I \LL [ Y_{IJ\ldots}]&=& \LL [ B^I Y_{IJ\ldots}] \nl
(T_B\lambda)^I  \LL [Y_{AI\ldots}]&=& \LL [ (T_B\lambda)^I Y_{AI\ldots}] \nl
\LL[U^I_J]&=& \gamma^I_{KL}B^KU_J^L\nl
\LL[\wt B^I ]&=& -\gamma^I_{JK }B^J\wt B^K
\eea
where $Y_{\ldots}$ stands for an arbitrary covariant function of $\lambda^I$, and $U_J^I$ is defined in \eqref{eq_anomalous_dimensions_app}.

\subsection{Gravitational terms and their  Weyl variations}
\bea
W^2&=&R^{\mu\nu\rho\sigma}R_{\mu\nu\rho\sigma}-2R^{\mu\nu}R_{\mu\nu}+\frac 1 3 R^2\nl
E_4&=&R^{\mu\nu\rho\sigma}R_{\mu\nu\rho\sigma}-4R^{\mu\nu}R_{\mu\nu}+R^2
\nl
G_{\mu\nu}&=&R_{\mu\nu}-\half  g_{\mu\nu}R
\eea
\bea
 \Delta^g_\sigma  g^{\mu\nu}&=&2\sigma  g^{\mu\nu}
\nl
\Delta^g_\sigma \sqrt {- g}&=&-4\sigma \sqrt {- g}\nl
\Delta^g_\sigma \nabla_\mu \nabla_\nu f
&=&
2\d_{(\mu}\sigma\d_{\nu)}f
- g_{\mu\nu}\d^\rho\sigma\d_\rho f
\nl
\Delta^g_\sigma\nabla^2 f
&=&2\sigma\nabla^2f
-2\d_\mu\sigma \d^\mu f
\nl
\Delta^g_\sigma R&=&2\sigma R
+6\nabla^2\sigma\nl
\Delta^g_\sigma G_{\mu\nu}&=&
2 \nabla_{(\mu}\nabla_{\nu)}\sigma
-2 g_{\mu\nu}\nabla^2\sigma\nl
\Delta^g_\sigma\sqrt {- g}W^2&=&0\nl
\Delta^g_\sigma\sqrt {- g}E_4
&=&-8\sqrt {- g}G^{\mu\nu}\nabla_\mu\nabla_\nu\sigma
\eea
\subsection{Weyl Variations of dimensionful functions of the sources}

\bea
\label{eq_useful_variations}
\Dw_\sigma\l Y_I\Lambda ^I \r
&=&
\sigma\l  2Y_I\Lambda ^I 
-\LL[ Y_I] \Lambda ^I
-Y_{I} \gamma^I_{JK}   \Pi^{JK}\r 
-2\n_\mu\sigma \l  Y_I\n^\mu \lambda^I\r 
\nl
\Dw_\sigma\l  Y_{IJ}\Pi^{IJ}\r  &=&
\sigma \l 2 Y_{IJ}\Pi^{IJ} - \LL[Y_{IJ}]\Pi^{IJ}
+ Y_{IJ}\gamma^{IJ}_{KL} \Pi^{KL}\r\nl
\Dw_\sigma (Y_a  \Pi^a) &=& \sigma\l  2Y_a \Pi^a - \LL[Y_a] \Pi^a  +\gamma_{IJ}^{~a} \Pi^{IJ}  \r\nl
\Dw_\sigma (Y_A F^A_{\mu\nu}) 
&=&\sigma\l - \LL[Y_A] F^A_{\mu\nu}- 2Y_A\d_{[J} P_{I]}^A\n_{\mu}\lambda^J\n_{\nu}\lambda^I \r 
-\n_{[\mu}\sigma\l 2 Y_A P_I^A\n_{\nu]}\lambda^I\r \nl
 \Dw_\sigma
   (Y_{IJK} \Omega^{IJK}) 
&=&
\sigma 
 \l 4Y_{IJK} \Omega^{IJK}-\LL[Y_{IJK}]\Omega^{IJK}+Y_{IJK}    \gamma^K_{MN}\Pi^{IJ}   \Pi^{MN}
\r \nl&&+   \n_\mu \sigma\l -2Y_{IJK} \n_\nu\lambda^I\n^\nu\lambda^J  \n^\mu \lambda^K\r 
-  B^I  Y_{I[JK]} \Xi_\sigma^{JK}
\eea
where the $Y$'s are arbitrary covariant functions of $\lambda^I$.

\section{Weyl symmetry in a regulated theory}
\label{app_CS_derivation}
In this appendix we shall give more details concerning the local CS equation. In particular we shall outline its derivation in dimensional regularization in weakly coupled 4D field theory and explicitly derive the structure of the anomaly and its consistency condition in 2D field theory.

First of all we want to explain how to find the Weyl transformation for the sources ${\cal J}$. An explicit way to do that is by a variant of the dilaton trick \cite{Luty:2012ww}. In order to see how that works, let us focus for the moment on a classical bare action $S^{(1)}[\Phi, g_{\mu\nu}, {\cal I}_0]$, where ${\cal I}_0$ indicates the general set of bare sources, the metric excluded, that can couple to non-trivial local functions of  $\Phi$ and of its derivatives. In the case of a theory regulated with a momentum cut-off such as Pauli-Villars one should add to the set  ${\cal I}_0$ also  the regulator mass $\Lambda$.
Now, the trick is to write the metric in a redundant way by introducing a dilaton field $\tau$: $S^{(1)}\equiv S^{(1)}[\Phi,e^{2\tau}g_{\mu\nu},{\cal I}_0]$. The action so written is trivially invariant under a Weyl transformation under which $\tau\to \tau +\sigma$, $g_{\mu\nu}\to g_{\mu\nu}e^{-2\sigma}$, while $\Phi$, ${\cal I}_0$ (and the regulator mass) do not transform. Now, if, and only if, ${\cal I}_0$ includes {\it all} the sources that can couple to the fields $\Phi$, we can certainly absorb $\tau$ in the fields and in the sources (and regulator if needed): $S^{(1)}[\Phi,e^{2\tau}g_{\mu\nu},{\cal I}_0]=S^{(1)}[\Phi^\tau,g_{\mu\nu},{\cal I}_0^\tau]$. Now, the redefined fields and sources, via their $\tau$ dependence, transform in a definite way under Weyl so as to compensate the transformation of the metric, and ensure formal invariance of the action. That is most easily understood by working around $\tau=0$ which gives
\beq
\delta_{\sigma} {\cal I}_0\equiv  {\cal I}_0^{\sigma} - {\cal I}_0\, .
\label{I_0}
\eeq
The situation is particularly neat when dimensional regularization (DR) can be used. In DR, the regulator itself is Weyl invariant and only the bare sources transform non-trivially. On the other hand, in the case of a momentum regulator, such as Pauli-Villars, things are a bit more involved as one must also consider a $\tau$ dependent, and consequent Weyl transforming, regulator mass: $\delta _\sigma\Lambda=\sigma \Lambda$. An obvious generalization of RG invariance then ensures that the combination of the transformation in  eq.~(\ref{I_0}) together with  $\delta _\sigma\Lambda=\sigma \Lambda$ has the same effect on the partition function as a certain transformation $\delta_\sigma {\cal I}$ of the renormalized sources ${\cal I}$. The latter combined with $\delta_\sigma g_{\mu\nu}=-2\sigma g_{\mu\nu}$, defines the transformation of renormalized sources ${\cal J}$. According to the discussion at the end of  section \ref{sec_setup}, the local Callan Symanzik equation then follows.
 
 Consider now a 4D renormalizable field theory based on a gauge group $G$, and involving  scalars and fermion transforming in a representation of $G$. In addition to the metric $g_{\mu\nu}$, the set of sources ${\cal J}$ consists of
 \begin{itemize}
\item the marginal couplings $\lambda^I\,\equiv \,$ gauge, Yukawa and scalar quartic couplings
 \item the gauge fields $A_\mu^A$ of the flavor symmetry group $G_F$ of the kinetic term; this symmetry is in general broken by the Yukawa and quartic couplings\footnote{As we assume our theory respects parity we just need to focus on the maximal vectorlike subgroup of $G_F$.}.
 \item mass terms $m^a$  for the scalar bilinears.
 \end{itemize}
 The general relation between the bare and the renormalized sources is obtained by considering all the  terms allowed by symmetry and power counting
\bea
\label{eq_bare_A}
\lambda^I_0(x)&=& \mu^{k^I\epsilon} \l \lambda^I(x,\mu) + L^I \r\nl
A_{0\mu}^A(x)&=& A_\mu^A(x,\mu) + N_I^A\n_\mu\lambda^I\nl
m_0^a(x)&=&\l \l \delta^a_b+ Z_{b}^a\r m^b(x,\mu)+ Z^a R(g)  + Z^a_I \n^2\lambda^I + Z^a_{IJ}\n_\mu\lambda^I \n^\mu \lambda^J\r
\eea
where $L^I,\,N_I^A,\,Z_{b}^a,\,Z^a,\,Z^a_I,\,Z^a_{IJ}$ are series of poles in $\epsilon$ whose coefficients are polynomial series in $\lambda$. The coefficients $k^I$ (understood not to be part of the summation convention) correspond to the dimensionality of the bare couplings in $4+\epsilon$. The $k^I$ equal $-1,\,-1,\, -1/2$ for respectively gauge, scalar quartic and Yukawa couplings. Notice that the dimensionality of $A_{0\mu}^A$ and $m_0^a$ is not affected by dimensional continuation. Notice also that the bare and the renormalized metric  can be taken to coincide. The effective action is renormalized by adding the most general set of diffeomorphism invariant counterterms:  these  can be absorbed in redefinitions of the fields and  sources in eq.~(\ref{eq_bare_A}), with no need  to redefine $g_{\mu\nu}$.  
By inspection of the most general dimensionally continued bare action $S^{(1)}$,  the Weyl transformation of the bare sources is simply given by
\beq
(g^{\mu\nu}, \,\lambda_0^I,\, A_{0\mu}^A,\, m_0^a)\quad\longrightarrow\quad (e^{2\sigma} g^{\mu\nu}, \,e^{k_I\epsilon \sigma} \lambda_0^I,\, A_{0\mu}^A,\, e^{2\sigma}m_0^a)
\label{baretransf}
\eeq
By eq.~(\ref{eq_bare_A}) this can be univocally translated into the, generally more involved, tranformation law for the renormalized sources
\beq
\delta_\sigma {\cal J}\equiv (2\sigma g^{\mu\nu}, \,\delta_\sigma \lambda^I,\,\delta_\sigma A_\mu^A,\,\delta_\sigma m^a)\, .
\eeq
In practice, the first of eqs.~(\ref{eq_bare_A}) fixes $\delta_\sigma \lambda^I$, and  once that is fixed  the second equation fixes $\delta_\sigma A_\mu^A$. Finally, once all other tranformations are fixed the the third equation can be used to deduce $\delta_\sigma m^a$.
By applying the logic described in section \ref{sec_setup}, we thus conclude the renormalized action must satisfy an equation of the form  
\beq
\label{eq_DR_CS_equation}
\int d^D x \l \delta_\sigma {\cal J}\frac{\delta}{\delta {\cal J}}
\r \WW\,=\, \int d^D x \delta_\sigma S^{(2)}[{\cal J}]\equiv \int d^D x {\cal A}_\sigma
\eeq
By this equation, given the finiteness of  $g^{\mu\nu}\frac{\delta}{\delta g^{\mu\nu}}{\cal W}$ and the finiteness of derivatives with respect to the renormalized sources, one deduces that  $(\delta_\sigma \lambda^I,\,\delta_\sigma A_\mu^A,\,\delta_\sigma m^a)$ and ${\cal A}(x)$ must also be finite. In other words: given $T$ is finite, then the coefficients of its expansion in terms of renormalized operators must be finite, along with the contact terms associated with the anomaly. The condition of $T$ finiteness is at the basis of the derivation of consistency conditions given in ref. \cite{Jack:1990eb}. Finiteness then allows us to safely take the $n\to 4$ limit in the above equation. This is the formal derivation of the local CS equation. In the following sections we shall describe in detail the structure of $(\delta_\sigma \lambda^I,\,\delta_\sigma A_\mu^A,\,\delta_\sigma m^a)$.

\subsection{The variation of $\lambda^I$}
$\delta_\sigma\lambda^I$  can be found using the following manipulation
\bea
e^{\sigma\epsilon k^I}   \lambda^I_0(x)&=&
  e^{\sigma\epsilon k^I} \mu^{k^I\epsilon} \l \lambda^I(x,\mu) + L^I \l\lambda(x,\mu), \epsilon\r\r\nl
&=&   \mu^{k^I\epsilon} \l \lambda^I(x,e^{-\sigma} \mu ) + L^I \l\lambda(x,e^{- \sigma} \mu), \epsilon\r\r
\eea
where we used the $\mu$ independence of the bare sources. 
In other words, a Weyl transformation for the bare sources is equivalent to a change in the renormalization scale:
\bea
  \lambda^I_0\to e^{\sigma\epsilon k^I} \lambda^I_0 \qquad \Longrightarrow \qquad
  \lambda^I (x,\mu)\to \lambda^I (x,e^{-\sigma}\mu)
\eea
In assigning these transformation properties it was essential that the sources are $x$ dependent by definition. This can be translated into the following infinitesimal transformation law for the renormalized sources
\bea
\delta_\sigma \lambda^I(x,\mu)=-\sigma(x)\frac {d}{d\log \mu}\lambda^I(x,\mu)
&\equiv&-\sigma \hat \beta^I 
\eea
In agreement with the local CS equation.

The last step is to relate the $\hat \beta$-function to the poles in the counterterm. This is done by using the invariance of the bare parameters under change of renormalization scale:
\bea
\label{eq_beta_function}
\mu\frac {d\lambda_0}{d\mu}=0  \Rightarrow \qquad
 \l \delta_J^I + \d_J L^I\r \mu\frac {d\lambda^J} {d\mu}&=&-\epsilon k^I \l \lambda^I  + L^I \r
\eea
Using the finiteness of $\lambda_I$ we find in the $\epsilon\to 0$ limit
\bea
\hat \beta^I&\to&\beta^I=-k^IL_1^I+k^J\lambda^J\d_J L^I_1
\eea

\subsection{The variation of $A_\mu^A$}
 Unlike $\lambda_0^I$, $A_{0\mu}^A$ is invariant under the local scale transformation. Using this in eq. (\ref{eq_bare_A}) we find
\bea
\l \delta^A_B + \l N^A_I  (T_B)^I_J \lambda^J\r \r 
 \delta_\sigma  A_\mu^B
&=&
\sigma \l \hat \beta^J \d_JN^A_I+ N^A_J\d_I\hat\beta^J \r \n_\mu \lambda^I + N^A_I\hat\beta^I \n_\mu \sigma  
\eea
and we can identify the functions $\rho$ and $S$ from the local CS equation:
\bea
\label{eq_rho_s}
\l \delta_B^A + \l N^A  T_B \lambda\r \r 
  \rho_I^A &=&-\hat \beta^J \d_JN^A_I- N^A_J\d_I\hat\beta^J\nl
  \l \delta_B^A + \l N^A  T_B \lambda\r \r 
 S^B&=&N^A_I\hat\beta^I~.
\eea
Focusing on the $\epsilon$ independent terms in these equations, and using the finiteness of the renormalized sources, we find
\bea
\label{eq_L_N_and_CS}
S^A&=& -k^I\lambda^I N_{I,1}^A\nl
\rho_I^A &=&k^J\l \lambda^J \d_JN^A_{I,1} +N^A_{I,1} \r\nl 
P_I^A &=&k^J \lambda^J \l \d_JN^A_{I,1}-\d_IN^A_{J,1}\r 
\eea
where $L_1^I$ and $N_{I,1}^A$ are the coefficients of the simple poles in $L^I$ and $N_{I}^A$.

Let us now derive the consistency condition $B\cdot P=0$. 
First, we multiply the first line of (\ref{eq_rho_s}) by $\hat B^I=\hat \beta^I - (  S^A T_A \lambda)^I$
\bea
\l \delta_B^A + \l N^A  T_B \lambda\r \r 
  \hat B^I  \rho_I^A &=&
 - \hat \beta^I\d_I \l N^A_J \hat \beta^J\r  + S^B\hat \beta^I  \d_I \l \delta_B^A+N^AT_B\lambda\r
 \eea
where we used the covariance of $\hat \beta$, namely $\l T \lambda\r^I\d_I \hat \beta^J=( T\hat \beta)^J$.
Next, we substitute the second line of eq. (\ref{eq_rho_s}) and find
\bea
\l \delta_B^A + \l N^A  T_B \lambda\r \r 
  \l \hat B^I  \rho_I^A + \hat \beta^I\d_I S\r =0~. 
\eea
We conclude that 
\bea
\hat B^I   P^A_I \equiv\hat B^I\l  \rho^A_I  + \d_IS^A\r =0 
\eea
where we used the covariance of $S^A$  to show that $\l S T \lambda\r^I\d_I S =0$ 
and hence $\hat B^I\d_IS^A=\hat \beta^I  \d_IS^A$.

\subsection{Dim 2 operators}

Once the Weyl tranformations of $g^{\mu\nu}$, $\lambda^I$ and $A_\mu^A$ are fixed the expression for the bare source
\bea
m_0^a&=& \l \l \delta^a_b+ Z_{b}^a\r  m^b(\mu)+ Z^a R  + Z^a_I \n^2\lambda^I + Z^a_{IJ}\n_\mu\lambda^I \n^\mu \lambda^J\r\, ,
\eea
as well as its Weyl  transformation equation $m_0^a \to e^{2\sigma} m_0^a$, fix the coefficients functions in $\delta_\sigma m^a$.
\bea
&(2\delta^a_c+Z^a_c)\br \gamma^c_b&=\LL[Z^a_b]\nl
&	(2\delta^a_b+Z^a_b)C^b&=\LL[Z^a]\nl
&	(2\delta^a_b+Z^a_b)D^b_I&=\LL[Z^a_I]\nl
&	(2\delta^a_b+Z^a_b)E^b_{IJ}&=2Z^a_K\partial_I\partial_J\hat{\beta}^K+2\LL[Z^a_{IJ}]\nl
&	(2\delta^a_b+Z^a_b)\theta^b_I&=-2Z^a_I-2Z^a_J\partial_I\hat{\beta}^J-2\hat{\beta}^JZ^a_{IJ}\nl
&	(2\delta^a_b+Z^a_b)\eta^b&=\hat{\beta}^IZ^a_I-6Z^a
\eea
(for brevity we have ignored the contributions in the transformation related to global symmetries).
From these expressions
it is possible to derive the remaining consistency conditions \eqref{eq_consistency_condition}.

\subsection{Consistency conditions for the anomaly coefficients}
As an example for the derivation of the consistency conditions for the anomaly coefficients
we present the computation for the 2d case where the anomaly is given by (see \cite{Osborn:1991gm}):
\bea
\frac {1}{\sqrt {-g}}\AA_\sigma&=&\sigma\l -\half \beta_\Phi R +\half \chi_{IJ}\n_\mu\lambda^I\n^\mu \lambda^J  \r +\n^\mu\sigma \l w_I \n^\mu\lambda^I\r
\eea
For simplicity we will ignore the contributions from dimensionful sources.
The coefficients in this anomaly satisfy the consistency condition
\bea
\label{eq_CC_2D}
\d_I \beta_\Phi - \chi_{IJ} \beta^J + \LL[ w_I]&=&0
\eea

In dimensional regularization this anomaly can be understood as 
the result of the non-invariance of the following counterterms in the effective action
\bea
\WW &\supset& \int d^Dy \sqrt{-g}\mu^{\epsilon}\l \half b R +\half  c_{IJ}\n_\mu\lambda^I \n^\mu\lambda^J\r 
\eea
where $b$ and $c_{IJ}$ are understood as a series of poles in $\epsilon=D-2$, where the finite part is assumed to vanish.

Defining the symmetry generator of the regulated theory as 
\bea
\Dw_\sigma &=&\int d^D x ~ \sigma(x)\l \frac{\delta}{\delta \tau(x)} - \hat \beta^I\frac{\delta}{\delta \lambda^I(x)}\r  
\eea
where $\hat \beta^I =-\epsilon\lambda^I  + \beta^I$, we find
\bea
&&\Dw_\sigma
\int d^Dy \sqrt{-g}\mu^{\epsilon}\l \half b R +\half c_{IJ}\n_\mu\lambda^I \n^\mu\lambda^J\r\nl
&&=\int \sqrt {-g} d^Dx \l \sigma \l -\half \hat\beta_\Phi + \half \hat \chi_{IJ}\n_\mu \lambda^I\n^\mu \lambda^J\r 
+ \n_\mu\sigma \l \hat w_I \n^\mu\lambda^I\r \r
\eea
where 
\bea
\hat \beta_\Phi &=& \hat \beta^K \d_K b - \epsilon b\nl
\hat \chi_{IJ} &=&-\LL_{\hat \beta} [ c_{IJ}]+ \epsilon c_{IJ}\nl
\hat w_I &=& - (1+\epsilon) \d_I  b- c_{IJ} \hat \beta^J~.
\eea
The finiteness of $T$ ensures
that these specific combinations are necessarily finite.
In other words, in the $\epsilon=0$ 
limit we find $\hat \beta_\Phi\to\beta_\Phi$, $\hat \chi_{IJ}\to\chi_{IJ}$ and 
$\hat w_I \to w_I$.
Moreover, these coefficients satisfy the relation
\bea
\d_I \hat \beta_\Phi -\hat  \chi_{IJ}\hat \beta^J + \LL_{\hat\beta } [\hat w_I]
&=&
\epsilon \l -\d_I\hat \beta_\Phi + \hat w_I\r 
\eea 
which, in the $\epsilon=0$ limit, gives eq. \eqref{eq_CC_2D}.

\section{Unitarity and anomalous dimensions of currents}
\label{app_p_at_CFT}

In this appendix we would like to study in more detail the scale and conformal transformations of the operators, eq. \eqref{eq_K_transformations}, at a conformal fixed point. In particular, we would like to distinguish the primary scalars operators from the descendants of the non-conserved currents.
\\ Let us suppose the background couplings $\lambda^I$ break the flavor group $G_F$ down to a subgroup $H$. Let us to parametrize the coset $G_F/H$  with  indices  $A=1,\dots,m$, while the remaining indices $A=m+1,\dots,{\mathrm {dim}}_{G_F}$ parametrize the generators of $H$. Using the notation $v_A^I \equiv (T_A \lambda^I)$, we thus have that for $A=1,\dots,m$,  $v_A^I\neq 0$ are  $m$ linearly independent vectors, while  $v_A^I = 0$ for $A>m$. In block matrix notation we can write 
\begin{equation}
v = \left(
\begin{matrix}
\hat{v} & a  \\
0 & 0 
\end{matrix}
\right)
\label{v}
\end{equation}
where  $\hat{v}$ is a $m\times m$ matrix. The rows of $v$ run over the indices $A$, while its  columns run over the indices $I=1,\dots,N$:  $v$ is a rectangular ${\mathrm {dim}}_{G_F}\times N$ matrix.  Since $v_A^I$  are $m$ linearly-independent vectors,  $\hat{v}$ can be taken invertible by a proper linear tranformation in  $I$-space.

The  anomalous dimension matrix  for $J^\mu_A$ is:
\begin{equation}
\label{eq:anom_dim}
\gamma_{A}^B = v_A^I P_I^B\, .
\end{equation}
By the properties of unitary representation of  the conformal group it must vanish for the conserved currents and take the form
\begin{equation}
\label{eq:anom_dim_matrix}
\gamma = \left(
\begin{matrix}
\hat{\gamma} & 0  \\
0 & 0 
\end{matrix}
\right)
\end{equation}
with $\hat{\gamma}$ a diagonal and strictly positive definite matrix (thus invertible) acting on the subspace of broken generators. Now, using eqs.~(\ref{v}-\ref{eq:anom_dim_matrix}) $P$ is constraned to have the form 
\begin{equation}
\label{eq:pmatrix}
P = \left(
\begin{matrix}
\hat{v}^{-1} (\hat{\gamma} - a b) & -\hat{v}^{-1} a p \\
b & p 
\end{matrix}
\right)
\end{equation}
with $b$ an $(N-m)\times m$ matrix and $p$ is an $(N-m)\times ({\mathrm{dim}}_{G_F}-  m)$  matrix. Notice that $P$ is a transposed rectangular matrix with respect to $v$: rows run over $I$ and colums over $A$.
We can now go to a basis in $I$ space such that $v$ and $P$ are block-diagonal: 
\begin{equation}
v \to v'=v S^{-1} = \left(
\begin{matrix}
1 & 0  \\
0 & 0 
\end{matrix}
\right)
\end{equation}
\begin{equation}
P \to P'=S P = \left(
\begin{matrix}
\hat{\gamma} & 0 \\
0 & p 
\end{matrix}
\right)
\end{equation}
\begin{equation}
S = \left(
\begin{matrix}
\hat{v} & a \\
-b \hat{\gamma}^{-1} \hat{v} & \l 1 - b \hat{\gamma}^{-1} a\r 
\end{matrix}
\right)
\end{equation}
In the new basis, by eq.~(\ref{eq_naive_WI}) the operators $O_I$ $I=,1,\dots,m$ are  the descendants of the broken currents $J^\mu_A$, $A=1,\dots, m$. On  the  broken generator subspace $P$ equals the anomalous dimension matrix $\hat\gamma$. Correspondingly eq. \eqref{eq_K_transformations} gives, as expected, $K^\mu {\cal O}_\alpha=-\hat \gamma_\alpha^\beta J_\beta^\mu$ for $\alpha,\beta=1,\dots,m$. However, as long as $p\not =0$, eq. \eqref{eq_K_transformations} also implies $K^\mu {\cal O}_I=\sum_{A>m} p_i^AJ_A^\mu \not 0$ for the supposedly primary operators described by $I>m$ (notice the sum is over the conserved curents). We thus expect $p$ should vanish. The proof comes by using unitarity as follows.

Let us consider the 2-point correlator of a scalar field and an unbroken current:
\begin{equation}
\langle J_A^\mu(p) O_I(-p) \rangle = f(p^2) p^\mu 
\end{equation}
The conservation of the current $p_\mu J_A^\mu(p)$ implies $f(p^2) p^2 = 0$, thus $f(p^2) = 0$.
\begin{equation}
\langle J_A^\mu(x) O_I(0) \rangle = 0
\end{equation}
If we act with a conformal transformation:
\begin{equation}
0 = \langle [ K^\nu, J_A^\mu(x)] O_I(0) \rangle + \langle  J_A^\mu(x)[ K^\nu, O_I(0)] \rangle = p_I^B \langle J_A^\mu(x) J_B^\nu(0) \rangle 
\end{equation}
where the $B$ runs only over the non-conserved currents, since otherwise the 2-point function vanishes.
In a unitary theory $\langle J_A^\mu(x) J_B^\nu(0)  \rangle$ is invertible, thus $p_I^B = 0$.  
\section{The consistency conditions for the Weyl anomaly}
\label{app_osborn_anomaly}

The most general parameterization of the Weyl anomaly given in eq. \eqref{eq_osborns_anomaly} can be reduced 
by a change of scheme. More specifically, the terms proportional to $d$, $U_I$, $V_{IJ}$, $\wt S_{(IJ)}$, $T_{IJK}$, $k_a$, and $j_{aI}$ can be eliminated by adding to the generating functional $\WW$ a local functional 
 \beq 
 {\cal F}_{\n^2 R}=\int d^4x \sqrt{g} \LL_{\n^2 R}
 \eeq
 with
\bea
\label{eq_local_term_n2r}
\LL_{\n^2 R}&=&
\l d + \half B^I U_I\r   \frac {R^2}{36}  
+U_I\frac{R}{6} \n^2\lambda^I  
+\half  V_{IJ}  \frac{R}{6}  \n_\mu\lambda^I \n^\mu\lambda^J +\hat m^a  k_a  \frac R 6
\nl
&&
+\frac 1 4  T_{IJK} \Pi^{IJ} \Lambda ^K
 +\half   j_{aI}  \Pi^a \Lambda^I 
+\frac 1 4  \l  \wt S_{(IJ)} +\half  T_{IJK}B^K
+\half j_{aI}\theta^a_J\r \Lambda ^I \Lambda^J
\eea
In addition to eliminating the mentioned terms, this operation also changes the remaining anomaly coefficients (the specific expression are not particularly illuminating). In the following equations we assume that all the coefficients are given in the scheme where these terms are indeed vanishing.

A key observation is that in this scheme the consistency conditions can be written as algebraic constraints. 
Here we list the equations, and the terms in the LHS of \eqref{WZ}  to which they are related:
\bea
\begin{array}{l l  l l ll }
\sigma_{[1} \n_\mu \sigma_{2]} \n^\mu R&~~: ~~&\beta_c &=&- \frac 1 4   \chi_{I}^e B^I 
 \nl\tabspace
\sigma_{[1} \n_\mu \sigma_{2]} \n^\mu \n^2\lambda^I
&~~: ~~ &\chi_I^e &=&- \half \chi_{IJ}^a B^J 
 \nl\tabspace
\n^2\sigma_{[1} \n_\mu \sigma_{2]} \n^\mu \lambda^I
&~~: ~~ &
 Y_I - \chi_I^e  &=&-\half  \wt S_{[IJ]}B^J 
 \nl\tabspace
\sigma_{[1}\n^2\sigma_{2]} \n_\mu \lambda^I \n^\mu \lambda^J
&~~: ~~ &
 \chi_{IJ}^f&=&
\half \chi_{IJ}^g+\half  \chi_{IJK}^bB^K- \d_{(J}\l \chi_{I)K}^aB^K\r
 \nl\tabspace
\sigma_{[1} \n^2 \sigma_{2]} \hat m^a
&~~: ~~ &q_a&=& \half r_{aI} B^I 
 \nl\tabspace
\sigma_{[1} \n_\mu \sigma_{2]} \n^\mu\lambda^I \hat m^a
&~~: ~~ &
 r_{aJ}U^J_I&=&   -\half  s_{aIJ} B^J-\half p_{ab}\theta_I^b
 \nl\tabspace
\sigma_{[1} \n_\mu \sigma_{2]} \n^\mu \lambda^{(I} \Lambda^{J)}
&~~: ~~ &\chi_{KL}^aU^K_IU^L_J&=&
\frac{1}{4}p_{ab}\theta_I^a\theta_J^b
+\half s_{a(JK}\theta^a_{I)}B^K
-\half  B^K  \chi_{K(IL}^bU^L_{J)}
-\half  \chi_{(IK}^gU^K_{J)}
 \nl\tabspace
\sigma_{[1} \n_\mu \sigma_{2]}\n^\mu \lambda^K \n_\nu \lambda^I\n^\nu \lambda^J 
&~~: ~~ &\chi_{IJL}^BU^L_K &=&-\half s_{aIJ}\theta^a_K+\br\chi_{IJK}^g-\chi_{IJKL}^cB^L + \wt S_{[KM]}\gamma^M_{IJ}
\nl\tabspace
& & &&-\half \left( \zeta_{AJK}P_I^A+\zeta_{AIK}P_J^A \right) -\left(\eta_{AJ}\d_{[K}P_{I]}^A+\eta_{AI}\d_{[K}P_{J]}^A \right)\nl\tabspace
  \n_{\mu} \sigma_{[1} \n_{\nu}\sigma_{2]}\n^\mu\lambda^{I}\n^\nu\lambda^{J}
&~~: ~~ &
 \wt S_{[IJ]}&=&\d_{[J}w_{I]}+\eta_{A[J} P^A_{I]}
\end{array}
\eea
The three non-trivial consistency conditions and the corresponding terms in the commutator are
\bea
\begin{array}{l l  l l ll }
\sigma_{[1} \n_\mu \sigma_{2]} G^{\mu\nu} \n_\nu \lambda^{I} 
&~~: ~~ &
\LL [ w_I]&=&-8\d_I\beta_b+\chi_{IJ}^gB^J\nl
 \tabspace
\sigma_{[1} \n^\mu \sigma_{2]} F^{A}_{\mu\nu} \n^\nu \lambda^{I} 
&~~: ~~ &
\LL [ \eta_{AI}]&=&\kappa_{AB}P^B_I+\zeta_{AIJ}B^J- \chi_{IJ}^g (T_A\lambda)^J\nl
 \tabspace
\n^\mu\sigma_{[1} \n^\nu \sigma_{2]} F^{A}_{\mu\nu} 
&~~: ~~ &
\eta_{AI}B^I&=&- w_I (T_A\lambda)^I  \tabspace
\end{array}
\eea
The coefficient of the last term in the commutator, $\sigma_{[1} \n_\mu \sigma_{2]} \n^\mu \lambda^{[I} \Lambda^{J]}$, vanishes by imposing the three unresolved consistency conditions, without introducing new constraints.

The anomaly coefficients appearing in \ref{sec_anomaly} are related to the ones appearing in the original formulation of the anomaly via
\bea
a&=&\beta_b,~~c=-\beta_a,\nl
b_{ab}&=&p_{ab}\nl
b_{aIJ}&=&s_{aIJ}   - j_{aK}\gamma^K_{IJ},\nl
b_{IJKL}&=&\chi_{IJKL}^c 
- \half T_{IJM}\gamma^M_{KL}
-\half  T_{KLM}\gamma^M_{IJ} 
~.\eea

\section{Computation of $\Gamma_{non-local}$ off-shell}
\label{app_non_local}

In this appendix we present a method for computing the non-local part of the dilaton effective action, without imposing the on-shell condition.
In \eqref{eq_dilaton_eft_exponent} we found an expression for this action which was obtained using the BCH formula
\bea
\label{eq_eft_non_local}
\Gamma_{non-local}[\gf,\tau]&= &
\exp\left \{\Delta_\tau^\beta+\half \left[ \Delta^g_\tau, \Delta^\beta_\tau-\Delta^g_\tau\right]+\ldots\right\}\WW\Big|_{\JJ_0}
\eea
To make use of this expression, it is necessary to bring it to a normal-ordered form, namely, to write the exponent in a form where all the derivatives are brought to the right. This can be done using the following useful relation:
\bea
\label{eq_normal_ordering}
\exp\left \{\mathcal F \frac{\delta} {\delta\JJ}\right \}&=&: \exp\left \{ \sum_{k=1}^{\infty}\frac{1}{k!}\mathcal F_k\frac{\delta} {\delta\JJ}\right \}:
\eea
where $:~:$ denotes normal ordering, $\mathcal F$ is some function of the sources $\JJ$ and $\mathcal F_k$ is the first order differential operator in $\l \mathcal F \frac{\delta} {\delta\JJ}\r ^k$ which can be defined recursively via 
\bea
\label{eq_recursion_for_F}
\mathcal F_k=\mathcal F \frac{\delta} {\delta\JJ}\mathcal  F_{k-1}~.
\eea
This relation, as well as the convenient transformation properties of the $\Pi$ functions given in \eqref{eq_pi_transformation}, will play an important role in the computation of the non-local part of the effective action.

\subsection{$\Gamma_{non-local}$ without dimension 2 operators} 
To demonstrate the method of our computation, we begin by considering a theory without dimension 2 operators, where all the commutators in \eqref{eq_eft_non_local} vanish and \eqref{eq_normal_ordering} can be used to write
\bea
\label{eq_eft_non_local_Delta_tau_k}
\Gamma_{non-local}[\tau]&= &
\exp\{ \Delta_\tau^\beta\}\WW= : \exp\left \{ \sum_{k=1}^{\infty}\frac{1}{k!} \Delta^\beta_{\tau,k}\right \}: \WW\big|_{\JJ_0}
\eea
where the most general form of $ \Delta^\beta_{\tau,k}$ can be parameterized via
\bea
\label{Delta_tau_k_no_dim2}
 \Delta_{\tau,k}^\beta&=&
\int d^4x \l 
\tau^k\l v^I_{,k}  \frac{\delta}{\delta \lambda^I(x)} + \l v^A_{I,k} \n_\mu\lambda^I \r  \frac{\delta}{\delta A^A_\mu(x)}\r 
+\tau^{k-1}\n_\mu\tau \l u^A_{,k} \frac{\delta}{\delta A^A_\mu(x)} \r \r ~.
  \nl
\eea
Using eq. \eqref{eq_recursion_for_F} one easily finds a recursive expression for the coefficients appearing in this formula:
\bea
\label{eq_recursion_v_k_I}
\begin{array}{lllllll}
v_{,1}^I&=&B^I &\qquad&v_{,k}^I&=& B^J\d_J v_{k-1}^I\\
 v_{I,1}^A&=&P_I^A &\qquad&v_{I,k}^A&=&B^J\d_J v_{I,k-1}^A + \gamma_I ^J v_{J,k-1}^A\\
u_{,1}^A&=&0 &\qquad&u_{,k}^A&=&B^J\d_J u_{k-1}^A+ B^I v_{I,k-1}^A \\
\end{array}
\eea

What is the physical interpretation of this expression? Using the definition of the effective action \eqref{eq_eft_non_local_Delta_tau_k} the coefficients $v^I_k$, $v^A_{I,k}$ and $v_{,k}^A$ can be understood as the coupling of the  composite operators $\left[\OO_I\right]$ and $\left [J^\mu_A\right]$ to $k$ dilatons.  
Notice that, as in the computation described in section \ref{dilaton_action},  the dilaton decouples from the currents $\left[ J^\mu_A\right]$ in the gauge $S^A=0$.
This is based on the following observations:  
First, $v_{I,k}^A$ does not contribute in the limit $\n\lambda=0$. Also, it is easy to show by induction, using the consistency condition $B^IP_I^A=0$,  that   
also $B^Iv_{I,k}^A=0$ for any $k$. Plugging this into the third line of \eqref{eq_recursion_v_k_I}, and using the choice $u_1^A=-S^A=0$,  we find 
\bea
u_k^A=0
\eea
and therefore, in this gauge, both terms in \eqref{Delta_tau_k_no_dim2} containing  $\frac{\delta}{\delta A}$ can be ignored.

We conclude that in the absence of dimension 2 operators the non-local part of the dilaton effective action can be written as
\bea
\label{eq_Jiu_Jitsu}
\Gamma_{non-local}[\tau]=
: \exp\left \{ \int d^4x \sum_{k=1}^\infty \frac {\tau^k}{k!}  \l B^J\d_J\r ^{k-1}B^I  \frac{\delta}{\delta\lambda^I(x)}\right \}: \WW\big|_{\JJ_0}
\eea
where, in agreement with the result quoted in the text, the series can be summed to give the following expression
\bea
\sum_{k=1}^\infty \frac {\tau^k}{k!} \l B^J\d_J\r ^{k-1}B^I=\lambda^I(\mu e^{\tau})-\lambda^I(\mu)~.
\eea

\subsection{$\Gamma_{non-local}$ in the presence of  dimension 2 operators} 
The computation of the effective action in the presence of dimension 2 operators is more complicated because the commutator in eq. \eqref{eq_eft_non_local} is not vanishing 
\bea
\label{eq_non_local_with_Oa}
\Gamma_{non-local}[\gf,\tau]&= &
\exp\left\{\Delta^g_\tau\right\} \exp\left\{-\Delta_\tau\right\}\WW\Big|_{\JJ_0}\nl
&=&
\exp\left \{\Delta_\tau^\beta-\half \left[ \Delta^g_\tau, \Delta_\tau\right]+\ldots\right\}\WW\Big|_{\JJ_0}\nl
&\equiv& \exp\{ \wt \Delta_\tau^\beta\}\WW\Big|_{\JJ_0}
\eea
We will now explain how to find the normal ordered form of the effective action
\bea
\label{eq_eft_non_local_tilde_Delta_tau_k}
\Gamma_{non-local}[\gf, \tau]&= &
\exp\{ \wt \Delta_\tau^\beta\}\WW\Big|_{\JJ_0}= : \exp\left \{ \sum_{k=1}^{\infty}\frac{1}{k!} \wt \Delta^\beta_{\tau,k}\right \}: \WW\Big|_{\JJ_0}
\eea
without using the BCH formula explicitly. As a first step, we rewrite the first order differential operator in the product of exponents appearing in \eqref{eq_non_local_with_Oa} as follows:
\bea
\label{eq_product_of_exponents}
\left [ \exp\left\{\Delta^g_\tau\right\} \exp\left\{-\Delta_\tau\right\}\right] _{1st-order,~\JJ_0}
&=&\sum_{n=0}^\infty \sum_{m=0}^\infty \frac{1}{m!n!}\left [ (\Delta^g_\tau)^n(-\Delta_\tau)^m\right] _{1st-order,~\JJ_0}\nl
&=&\left [ \Delta^\beta_\tau  + \sum_{k=2}^\infty \frac{1}{k!}(-\Delta_\tau)^k\right] _{1st-order, ~\JJ_1}
\eea
where the notation $\left [ \ldots \right] _{1st-order}$ stands for keeping only the first order differential operator, and the sources $\JJ_0$ and $\JJ_1$ were defined in eqs. \eqref{eq_J_1} and \eqref{eq_J_0}.
Comparing this expression with \eqref{eq_eft_non_local_tilde_Delta_tau_k}, and matching the first order differential operators in both expressions, we find for $k>1$:
\bea
\wt\Delta_{\tau,{k}}^\beta
&=& 
 \left[(-\Delta_\tau)^k\right]_{1st-order,\JJ_1}=
(-1)^k \Delta_{\tau,k}  \Big| _{\JJ_1}
\eea
where $\Delta_{\tau,k}$ is defined as in eq. \eqref{eq_recursion_for_F}. 

Next, we compute  $\Delta_{\tau,k}$ in the scheme $\theta_I^a=\eta^a=0$ where $m^a=\Pi^a$. 
The transformation properties of the $\Pi$ functions given in \eqref{eq_pi_transformation}, 
suggests to write the term in $\Delta_{\tau,k}$ which is proportional to $\frac{\delta}{\delta m}$ as
\bea
\label{eq_tilde_k_beta}
(-1)^k\Delta_{\tau,k}&\supset&
\int d^4x~ \tau^k \l v_{b,k}^a \Pi^b +v^a_{IJ,{k}}\Pi^{IJ}\r 
\frac{\delta}{\delta m^a(x)}
\eea
where the coefficients are defined recursively by
\beq
\label{eq_recursion_relation_v_a_b}
\begin{array}{lllllll}
v_{b,0}^a&=&\delta_b^a &\qquad&v_{b,k}^a&=&\bar \LL [ v_{b,k-1}^a] -2 v_{b,k-1}^a \\
 v_{IJ,0}^a&=&0 &\qquad&v_{IJ,k}^a&=&\bar \LL [ v_{IJ,k-1}^a]  -2 v_{IJ,k-1}^a -  \gamma^{KL}_{IJ} v_{KL,k-1}^a
 -v^a_{b,k-1}\gamma^b_{IJ}\\
\end{array}
\eeq
and the notation $\bar  \LL$ stands for the Lie derivative evaluated when ignoring the upper $a$ indices.
In the limit  $\n\lambda=A=0$ we can substitute $\Pi^{IJ}=-\frac{1}{6}B^{I}\wt B^J  R$ and use the consistency condition 
$
C^a=-\frac{1}{12}E_{IJ}^a B^I \wt   B^J  
$ to further simply this expression:\footnote{It is also necessary to use the identity $\gamma^{K}_{IJ}  B^I (U^{-1})^J_L 
=- \LL [(U^{-1})^K_L]   $. }
\bea
\label{eq_almost_Gamma_non_local}
\sum_{k=1}^\infty \frac{(-1)^k}{k!}\Delta_{\tau,k}&\supset&
\int d^4xe^{-2\tau}\sum_{k=1}^\infty \frac{(-1)^k}{k!}  \tau^k \l  \tilde  v_{b,k}^a m^b +\frac 1 6 \tilde  v^a_{,k} R\r 
\frac{\delta}{\delta m^a(x)}
\eea
where we also extracted a factor of $e^{-2\tau}$ and 
defined
\bea
\begin{array}{lllllll}
\tilde v_{b,0}^a&=&\delta_b^a &\qquad&\tilde v_{b,k}^a&=&B^I\d_I \tilde v_{b,k-1}^a + \gamma^c_b  \tilde v_{c,k-1}^a  \\
\tilde  v_{,0}^a&=&0 &\qquad&\tilde v_{,k}^a&=&B^I\d_I\tilde v_{,k-1}^a+6\tilde v^a_{b,k-1}C^b~.\\
\end{array}
\eea

Plugging these results into \eqref{eq_eft_non_local_tilde_Delta_tau_k} we finally find that the non-local part of the dilaton effective action, associated with the dimension 2 operators in this scheme, is given by
\bea
\label{eq_non_local_Gamma}
\Gamma_{non-local}[\gf,\tau]
\supset: \exp\left \{ \int d^4x \sum_{k=1}^{\infty}\frac{ \tilde v^a_{,k} }{k!} \tau^k \l  \n^2\tau - (\n\tau)^2\r   \frac{\delta}{\delta m^a(x)}
\right \}: \WW\Big|_{\JJ_0}
\eea

Next, let us introduce a non-zero $\eta^a$ and $\theta_I^a$. In this case, it is convenient to write the variation of $m^a$ as
\bea
\Delta_\sigma m^a&=& \sigma \l \l 2\delta^a_b-\gamma^a_b \r \Pi^b  +\half \gamma^a_{IJ} \Pi^{IJ}\r 
+
\Delta_\sigma\l \eta^b \frac  R 6 + \half \theta_I^b\Lambda^I\r 
\eea
Consequently, the general form of $\Delta_{\tau,k}$ can be factorized into two terms:
\bea
\label{eq_factorized_Delta_k}
(-1)^k\Delta_{\tau,k}&\supset&
\int d^4x~ \tau^k \l v_{b,k}^a \Pi^b+ v^a_{IJ,{k}}\Pi^{IJ} \r 
\frac{\delta}{\delta m^a(x)}\nl
&&+
\left [(-\Delta_\tau)^{k}  \l \eta^a \frac  R 6 + \half \theta_I^a\Lambda^I\r\right] \frac{\delta}{\delta m^a(x)}
\eea
The first line generates terms similar to eq. \eqref{eq_non_local_Gamma}, with a few modifications due to the appearance of $\eta^a$ and $\theta_I^a$ in the definition of $\Pi^a$ and in the consistency conditions:
\bea
\Gamma_{non-local}[\gf,\tau]
\supset: \exp\left \{ \int d^4x \sum_{k=1}^{\infty}\frac{\tau^k }{k!}\l \tilde v^a_{,k} -\tilde v_{b,k}^a \wt \eta^b\r    \l  \n^2\tau - (\n\tau)^2\r   \frac{\delta}{\delta m^a(x)}
\right \}: \WW\Big|_{\JJ_0}\nl
\eea
where $\wt\eta^a$ was defined in \eqref{eq_tilde_B_eta} and  
\bea
\label{eq_tilde_v}
\begin{array}{lllllll}
\tilde v_{b,0}^a&=&\delta_b^a &\qquad&\tilde v_{b,k}^a&=&B^I\d_I \tilde v_{b,k-1}^a + \gamma^c_b  \tilde v_{c,k-1}^a  \\
\tilde  v_{,0}^a&=&0 &\qquad&\tilde v_{,k}^a&=&B^I\d_I\tilde v_{,k-1}^a+\tilde v^a_{b,k-1}(6C^b + \LL[\tilde \eta^b])~.\\
\end{array}
\eea
Plugging the second line of \eqref{eq_factorized_Delta_k} into \eqref{eq_product_of_exponents}, one finds the following expression
\bea
&&\sum_{k=1}^\infty \frac{1}{k!}(-\Delta_\tau)^{k}  \l  \eta^a \frac  R 6 + \half \theta_I^a\Lambda^I\r\Big| _{\JJ_1,\gf=\eta} \nl
&&= -e^{2\tau}\tilde \eta_a \l \n^2\tau -(\n\tau)^2\r 
+\sum_{k=0}^\infty \frac{\tau^k}{k!}\l  (B^J\d_J)^{k}(\theta_I^a\wt B^I) \l  \n^2\tau - (\n\tau)^2\r 
+  (B^J\d_J)^{k}(\theta_I^a B^I)(\n\tau)^2  \r \nl
\eea

In conclusion, the non-local part of the dilaton effective action, which is a generalization of eq. \eqref{lefftau} to include off-shell dilatons,  
given as a series expansion in powers of $\tau$ is
\bea
\LL_{eff}&=&
\sum_{k=1}^{\infty}\frac{\tau^k}{k!}
\l \l B^J\d_J\r ^{k-1}B^I  \OO_I + \l \tilde v^a_{,k} -\tilde v_{b,k}^a \wt \eta^b\r\l  \n^2\tau - (\n\tau)^2\r \OO_a  \r 
\nl
&& +\sum_{k=0}^\infty \frac{\tau^k}{k!}\l  (B^J\d_J)^{k}(\theta_I^a\wt B^I) \l  \n^2\tau - (\n\tau)^2\r 
+  (B^J\d_J)^{k}(\theta_I^a B^I)(\n\tau)^2  \r \OO_a\nl
&&-\tilde \eta_ae^{2\tau}\l  \n^2\tau - (\n\tau)^2\r   \OO_a
\eea
where the coefficients $\tilde v^a_{,k}$ and $\tilde v_{b,k}^a$
are given in eq. \eqref{eq_tilde_v}.



\end{document}